\documentclass[a4paper,11pt]{article}
\usepackage[utf8]{inputenc}
\usepackage[T1]{fontenc}
\usepackage[english]{babel}
\usepackage[width=175mm,height=257mm]{geometry}
\usepackage{amsmath,amsfonts,amssymb,bm}
\usepackage{lmodern,graphicx,float,indentfirst,multirow,rotating,csquotes,cite,xcolor}
\usepackage[squaren]{SIunits}
\usepackage[textfont={footnotesize},labelfont={color=blue,bf,sf},labelsep=endash]{caption}
\usepackage[position=top,labelfont={color=blue,bf,sf}]{subfig}
\usepackage[colorlinks,citecolor=blue]{hyperref}

\begin{document}
\begin{titlepage}
\setcounter{page}{1}
\renewcommand{\thefootnote}{\fnsymbol{footnote}}
\begin{flushright}
%ucd-tpg 10-01\\
%arXiv:yymm.xxxx
\end{flushright}
\vspace{5mm}
\begin{center}
{\Large \bf {Zitterbewegung Effect in Graphene with Spacially \\ Modulated Potential}}

\vspace{5mm}

{\bf Abdellatif Kamal}$^{a}$
and
{\bf Ahmed Jellal\footnote{\sf %ajellal@ictp.it --
a.jellal@ucd.ac.ma}}$^{a,b}$ 

\vspace{5mm}

{$^{a}$\em Laboratory of Theoretical Physics,  %Department of Physics,
Faculty of Sciences, Choua\"ib Doukkali University},\\
{\em PO Box 20, 24000 El Jadida, Morocco}

{$^{b}$\em Canadian Quantum  Research Center,  
204-3002 32 Ave Vernon, \\ BC V1T 2L7,  Canada}

\vspace{30mm}

\begin{abstract}
    The Zitterbewegung (ZB) effect is investigated in  graphene 
    with spacially modulated potential near the original Dirac point (ODP) 
    and extra Dirac points (EDPs). Our calculations show that to get  the large ZB oscillations, the wave packet center must be at the angle $\theta_0=0$ for EDPs located at zero-energy, or $\theta_0=\pi/2$ for both ODP
    and EDPs  
    at finite energy $\varepsilon=m\pi$ ($m$ integer). By varying the parameters ($q_2, \mathbb{V}$) of the periodic potential  and the initial momentum ($\kappa_0, \theta_0$) of Gaussian wave packet, it is found that the frequency of the ZB oscillations is in the range $[10^{7}~\hertz, 10^{13}~\hertz$] depending on what type of EDP is generated and the amplitude  reaches hundreds of angstroms but their attenuation becomes very slow. 
    More analysis of the frequency 
    shows the  possibilities in experimentally realizing 
    the ZB effect in our system.
   
\end{abstract}

\end{center}
    \vspace{3cm}
{\bf Pacs}: {03.65.Pm, 72.80.Vp, 73.21.Cd, 03.65.Sq}\\
{\bf Keywords}: {Graphene, spacially modulated potential,  extra Dirac points, Zitterbewegung effect, frequency of oscillations.}

\end{titlepage}
%\hypersetup{pageanchor=true}
%\newpage
%==================================
\section{Introduction}\label{Intro}
%==================================

   Graphene was discovered in 2004 \cite{Novo} and is a single layer of carbon atoms arranged hexagonally in the honeycomb lattice. It is considered to be the miracle material of the future because of its remarkably properties. Indeed, graphene is
   flexible, transparent, extremely robust, can assume different electrical properties and the highest thermal conductivity of all known materials \cite{Castro}. This makes it extremely interesting for
potential industrial applications 
such as the filtration of sea water, paint, tires, internet of things, aircraft structures $\cdots$. 
It is also a laboratory of experimentally testing 
different effects 
such as the quantum Hall effect
and weak-localization \cite{Novoselov2005Nature}. 
It also allowed for the observation
of some subtle effects, previously accessible only to high energy
physics, like Klein tunneling and vacuum breakdown \cite{Katsnelson2007}. 
In contrary, it is difficult  to observe the Zitterbewegung (ZB) effect \cite{Schrodinger1930}   
in graphene 
because of
its high frequency and low amplitude. However, it could  be possible   
%observed
under current experiment
conditions when graphene is subject to 
 a periodic potential \cite{Wang2014PRA}.

The ZB effect is a  
rapid motion of free particles, which was 
originally proposed for relativistic quantum dynamics in 1930 by
Schrödinger  \cite{Schrodinger1930}. The
ZB effect is a high frequency (trembling motion)
of a particle in vacuum  resulting from the 
interference between the positive and negative energy
states of the relativistic Dirac particle.
Recently, the ZB effect has attracted more attention because numerous theoretical work was done particularly on graphene \cite{Novoselov2005Nature,Rusin2007PRB,Maksimova2008PRB,Romera2009PRB,Chaves2010PRB,Cserti2006PRB}, graphene superlattices \cite{Wang2014PRA,Deng2015PRB,Jianli2018JPCM}, graphene nanoribbons \cite{Ghosh2015PRB}, graphene quantum dots \cite{Garcia2014PRB}. According to 
\cite{Wang2014PRA} some conditions need to be considered in order 
to observe the ZB effect in graphene. Indeed, its oscillations should be sufficiently maintained over time, their frequency should be smaller than $10^{15}~\hertz$ and their amplitude must be higher than a few angstroms.

On the other hand, one of the most important features %influences 
of the periodic potential is that it can decrease the group velocity of fermions near Dirac points. More precisely, we have seen that in  graphene with spacially modulated potential  the 
group velocity shows different behavior according to nature of the Dirac point \cite{kamal2018EPJB}. It is found that
its ${v_y}$-component  near extra Dirac points (EDPs) at finite energy including original Dirac point (ODP)  decreases remarkably but %its $x$-direction group velocity 
the ${v_x}$-component  does not change.
%Differently, near EDPs at zero-energy in the case where  $q_2=0$, both components decrease. %$v_x$ and $v_y$ both of them decrease. %\cite{kamal2018EPJB}.
%
%We have studied the electronic band structures of massless Dirac fermions in symmetrical graphene superlattice with cells of three regions \cite{kamal2018EPJB}. We have determined the group velocities expressions near extra Dirac points (EDPs) which has numerically analyzed and showed that there is a highly anisotropy of group velocity. 
%However, both $x$- and $y$-direction group velocities will decrease for EDPs located at zero-energy, but $x$-direction contain unchanged group velocity near original Dirac point and EDPs located at finite energy $\varepsilon=m\pi$. 
Such anisotropy of the group velocity is actually depending on the applied  potential and therefore 
it is relevant to investigate its influence on
the ZB effect.

We study the ZB effect of fermions in graphene with a spacially 
modulated potential near ODP and EDPs. Our  system is a symmetrical graphene superlattice with cells of three regions \cite{kamal2018EPJB} and two regions are separated by another one of distance $q_2$. We use a method based on the implicit function
to 
explicitly determine the dispersion relation close to a given Dirac point 
%$(kDx , kDy , εD )$ %, we introduce
in terms of the group velocity components. Subsequently, we %introduce
consider the Heisenberg dynamics and show that 
 the time evolution of the  position operators are functions of different physical parameters. 
 %get 
 %the their first matrix elements as function of the physical parameters.
%By considering a Gaussian wave packet, 
By assuming that
the initial state of Dirac fermions can be described by a
two-dimensional wave packet, we show that 
the averages of position operators are depending on the frequency
of the ZB oscillations.
%
%In order to make the Zitterbewegung more approachable experimentally, we consider in this work the motion of massless Dirac fermions described by a Gaussian wave packet subjected to multi-unit graphene superlattice. We will study the Zitterbewegung effect in the vicinity of extra Dirac points by analyzing the oscillations of the norm of the position operator average, it is shown that, 
Later on, under various conditions of the strength of the potential,  distance $q_2$ as well others, %initial momentum of Gaussian wave packet, 
we numerically analyze
the %can be used to tun 
ZB oscillations. In particular, we show that 
the   frequency %of the ZB oscillations 
belongs to the interval $[10^{7}~\hertz, 10^{13}~\hertz$] depending on the nature of EDP, %what type of extra Dirac point is generated and 
the amplitude  reaches hundreds of angstroms and the attenuation becomes very slow.
Our results tell us that the ZB effect can be
controlled by  $q_2$ of the central region together with the potential height.

The  paper is organized as follows. In section \ref{Model}, we give general formulation of the problem for massless Dirac fermions in graphene with spacially modulated potential.  We investigate ZB effect near different ODP and EDPs by considering the time evolution of the position operators in section \ref{ZB}.
%extra Dirac points of massless Dirac fermions described by a Gaussian wave packet with finite momentum. 
In section \ref{Results}, we discuss the ZB oscillations under suitable conditions  and make comparison
with literature to show their relevance. We conclude our results in the final section.

%======================================
\section{Model and method}\label{Model}
%======================================

\begin{figure}[!htb]
     \centering
     \includegraphics[width=10cm, height=5cm]{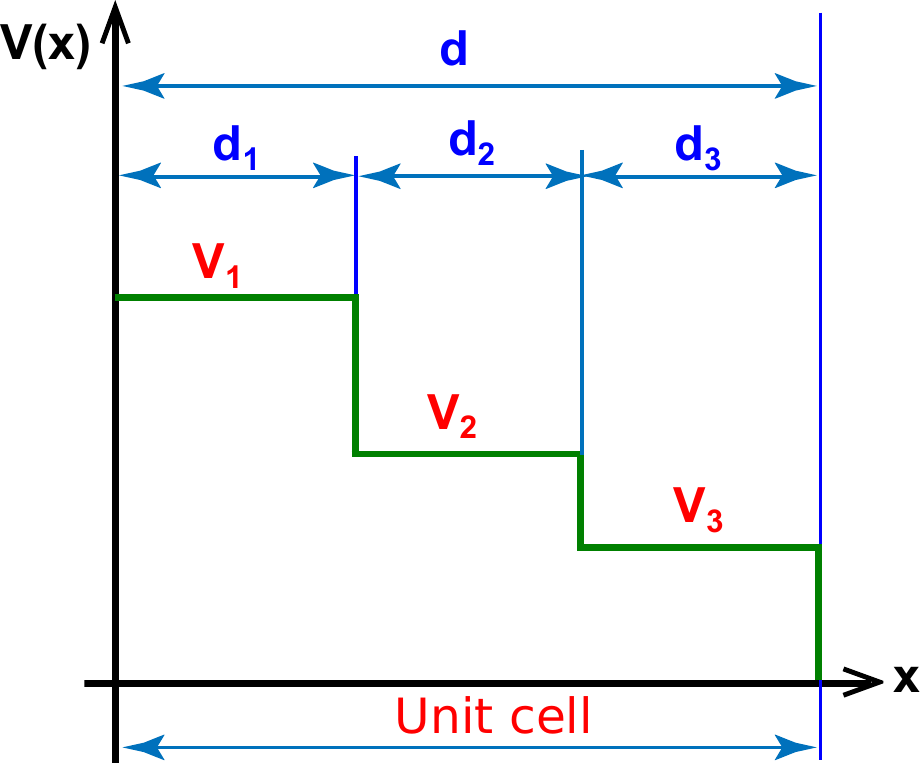}
     \caption{
        (color online) Schematic of the superlattice potential $V(x)$ composed of three regions growing along the $x$-direction with the period $d=d_1+d_2+d_3$, $d$  is the width of region $i$ and $V_i$ is its applied potential height.}
     \label{fig:model}
 \end{figure}

We propose another alternative way to study the ZB effect in graphene with spacially
 modulated potential near ODP and EDPs. Our approach is different to that used to deal with transient ZB in graphene superlattice \cite{Wang2014PRA}. For this,
let us consider  one dimensional periodic potential $\mathit{V(x)}$ composed of three regions growing along the $x$-direction with the period $d=d_1+d_2+d_3$.
This potential is applied to graphene with the height $V_i$ and distance $d_i$ 
of each region $i$, as depicted in 
%, $d_i$ is the width
%of region $i$, 
%see 
Figure \ref{fig:model}. %,  and applied to graphene with amplitude $V_i$ in region $i$.
The Hamiltonian %of the low-energy quasiparticles 
%in each 
describing the region $i$ of the $j^{th}$ elementary cell %(Figure \ref{fig:model}) 
can be written as
\begin{equation}
   H=\hbar v_F (-i\sigma_x \partial_x - i\sigma_y \partial_y)+V(x)\mathbb{I}.
   %v_F\bm{\sigma}\cdot\bm{p}+V(x)\mathbb{I}
    \label{eq:H_0}
\end{equation}
 %where $\bm{p}=(p_x,p_y)$ is the momentum operator, $v_F\approx 10^6 \meter\per\second$ is the Fermi velocity, $\bm{\sigma}=(\sigma_x,\sigma_y)$ are the Pauli matrices and $\mathbb{I}$ is the $2\times2$ unit matrix. First of all, 
  For convenience, we introduce dimensionless quantities %the distance 
  $q_i=d_i/d$ %with $i=1,2,3$ and 
  ($0\leq q_i\leq 1$), %giving rise $q=\left\{q_1,q_2,q_3\right\}$, the potential 
  $\mathbb{V}_i=V_i/E_F$ and %the energy 
  $\varepsilon=E/E_F$, with $E_F=\hbar v_F/d$ and $i=1,2,3$. %Each region $i$ of $j^{th}$ elementary cell (Figure \ref{fig:model}) has the Hamiltonian
 %and $G_0=2\pi/\mathit{d}$. The Hamiltonian $H_0$ then reads
%\begin{equation}
%    H_1=v_F\bm{\sigma}\cdot\bm{p}+V(x)\mathbb{I}
%    \label{Ham}
%\end{equation}
%where $\mathbb{I}$ is the $2\times2$ unit matrix.
 %where $V(x)$ is piecewise constant potential.
 We have already  studied the electronic band structures 
 %of massless Dirac fermions in symmetrical graphene superlattice with cells of three regions 
 corresponding to the Hamiltonian by determining the equation governing the dispersion relation \cite{kamal2018EPJB}. 
 
 We return back to our work \cite{kamal2018EPJB} and review one relevant part 
 to achieve our goals. Indeed,
to explicitly determine the dispersion relation close to a given Dirac point ($k_{D_x},k_{D_y},\varepsilon_D$), %we expand \eqref{33} around it. To do this, 
%we need to write this dispersion relation as an implicit function $f(k_x ,k_y,\varepsilon )=0$ with
we introduce the implicit function $f(k_x ,k_y,\varepsilon )=0$ with
\begin{eqnarray}\label{implicit}
  f(k_x ,k_y,\varepsilon )&=&\cos(k_x d)-\cos(k_1d_1)\cos(k_2d_2)\cos(k_3d_3)+G_{12}\sin(k_1d_1)\sin(k_2d_2)\cos(k_3d_3)\nonumber\\
    &&+G_{13}\sin(k_1d_1)\sin(k_3d_3)\cos(k_2d_2)+G_{23}\sin(k_2d_2)\sin(k_3d_3)\cos(k_1d_1)
    %= 0
\end{eqnarray}
%where we have set the quantities
and the parameters $G_{ij}$ are functions of the potential height and wave vectors
%such that the parameters are
\begin{equation}%\label{Gij}
      G_{ij}=\frac{(\mathbb{V}_i-\mathbb{V}_j)^2-(k_i^2+k_j^2)d^2}{2k_ik_jd^2},\qquad k_i=\frac{1}{d}\sqrt{(\varepsilon-\mathbb{V}_i)^2-(k_yd)^2}.
\end{equation}
At the Dirac point ($k_{D_x},k_{D_y},\varepsilon_D$), the  band structures are intersected. Then the gradient of dispersion relation must be equal zero \cite{Lima2015PLA379}, namely
\begin{equation}\label{grad}
  \bm{\nabla} f(k_{D_x},k_{D_y},\varepsilon_D)=0.
\end{equation}
%Now, we consider the Taylor approximation of $f$ near the Dirac point ($k_{D_x},k_{D_y},\varepsilon_D$) to write
Making the Taylor expansion of \eqref{implicit} around  ($k_{D_x},k_{D_y},\varepsilon_D$), we obtain the approximate form %\eqref{implicit} by
\begin{equation}\label{Taylor}
  f\left(k_x,k_y,\varepsilon\right)\approx f\left(k_{D_x},k_{D_y},\varepsilon_D\right)+\mathrm{\Delta}P\bm{\nabla} f\left(k_{D_x},k_{D_y},\varepsilon_D\right)+\frac{1}{2}\mathrm{\Delta}P^{t}
  \mathbb{H}f\left(k_{D_x},k_{D_y},\varepsilon_D\right)\mathrm{\Delta}P
\end{equation}
where the variation $\mathrm{\Delta}P$ and  Hessian matrix $\mathbb{H}$ are given by
\begin{equation}\label{GHessian}
    \mathrm{\Delta}P=\left(\begin{array}{c}
            \mathrm{\Delta}k_x \\
            \mathrm{\Delta}k_y \\
            \mathrm{\Delta}\varepsilon
          \end{array}\right)=\left(\begin{array}{c}
            k_x-k_{D_x} \\
            k_y-k_{D_y} \\
            \varepsilon-\varepsilon_D
          \end{array}\right),\qquad \mathbb{H}f\left(k_{D_x},k_{D_y},\varepsilon_D\right)=\left(
     \begin{array}{ccc}
       A & 0 & 0 \\
       0 & B & 0 \\
       0 & 0 & C \\
     \end{array}
   \right)
\end{equation}
%and the Hessian matrix $\mathbb{H}$ is a square matrix of second-order partial derivatives of $f$, which can be explicitly determined by fixing the nature of Dirac point. All the cases of Dirac points can be written in general form as
% \begin{equation}\label{GHessian}
%   \bm{\nabla} f\left(k_{D_x},k_{D_y},\varepsilon_D\right)=0, \qquad \mathbb{H}f\left(k_{D_x},k_{D_y},\varepsilon_D\right)=\left(
%      \begin{array}{ccc}
%        A & 0 & 0 \\
%        0 & B & 0 \\
%        0 & 0 & C \\
%      \end{array}
%    \right)
% \end{equation}
such that the parameters $A$, $B$ and $C$ are functions of %can be obtained in terms of  
the set $(q_i, \mathbb{V}_i, d_i)$ for a given Dirac point. Now injecting 
\eqref{grad} and \eqref{GHessian} into \eqref{Taylor} and using 
\eqref{implicit} to end up with 
%, \eqref{grad} and \eqref{GHessian} we write \eqref{Taylor} as 
\begin{equation}\label{quadratic}
 A\mathrm{\Delta}k_x^2+B \mathrm{\Delta}k_y^2=-C \mathrm{\Delta}\varepsilon^2
\end{equation}
%or equivalently
giving rise to the dispersion relation near the Dirac point ($k_{D_x},k_{D_y},\varepsilon_D$)
\begin{equation}\label{Diracdispersion}
  \varepsilon=\pm \sqrt{-\frac{A}{C}\left(k_x-k_{D_x}\right)^2-\frac{B}{C}\left(k_y-k_{D_y}\right)^2}+\varepsilon_D
\end{equation}
where sign($A$) = sign($B$) = $-$sign($C$) and the energy $\varepsilon_D$ 
%at Dirac point 
will be determined in in the next. 
%This is exactly the dispersion relation near the Dirac point and describes the anisotropy of energy contours.
%
%On the other hand, 
It is clearly see that  \eqref{Diracdispersion} is depending on % , we determine 
the group velocity associated to each Dirac point, such that
its the two normalized components are %of the normalized
%, which are %. They are defined by
\begin{equation}\label{GV}
u_x=  \frac{v_x}{v_F}=\frac{1}{d}\frac{\partial\varepsilon}{\partial k_x}, \qquad 
u_y=\frac{v_y}{v_F}=\frac{1}{d}\frac{\partial\varepsilon}{\partial k_y}.
\end{equation}
%In the next, we set the normalized group velocity components as 
%$\alpha=\frac{v_x}{v_F}$ and $\beta=\frac{v_y}{v_F}$.
Near Dirac points, we can approximate \eqref{GV}  using \eqref{quadratic} to obtain
the two components
\begin{equation}\label{GV3}
 % \frac{v_x}{v_F}
  u_x=\frac{1}{d}\left(\frac{\mathrm{\Delta}\varepsilon}{\mathrm{\Delta} k_x}\right)_{\mathrm{\Delta} k_y=0}=\frac{1}{d}\sqrt{-\frac{A}{C}}, \qquad %\frac{v_y}{v_F}
  u_y =\frac{1}{d}\left(\frac{\mathrm{\Delta}\varepsilon}{\mathrm{\Delta} k_y}\right)_{\mathrm{\Delta} k_x=0}=\frac{1}{d}\sqrt{-\frac{B}{C}}.
\end{equation}
%Finally, using \eqref{GV3} 
These can be implemented in
%which allows us to write 
\eqref{Diracdispersion} to get another form of the energy in terms of the wave vector
\begin{equation}\label{Diracdispersionyyy}
  \varepsilon=\pm   \kappa_e d+\varepsilon_D, \qquad 
%\end{equation}
%where %we have set 
%the wave vector is given by
%\begin{equation}\label{1200}
   \kappa_e=\sqrt{u_x^2\left(k_x-k_{D_x}\right)^2+
u_y^2\left(k_y-k_{D_y}\right)^2}.
\end{equation}
and therefore we can immediately realize that 
the corresponding
Hamiltonian takes the form
\begin{equation}
  H=\hbar v_F\left[u_x \left(k_x-k_{D_x}\right) \sigma_x+u_y \left(k_y-k_{D_y}\right) \sigma_y\right]+\varepsilon_D E_F \mathbb{I}.
  \label{eq:M_prime}
\end{equation}
This is actually different to the Hamiltonian obtained under some approximations by studying 
new generation of massless Dirac fermions in graphene under
external periodic potential %where some approximation are used
\cite{Park2008PRL}.
%
%after making different
%transormations.
To complete the above derivation, 
we calculate $\varepsilon_D$, %explicitly determine the energy 
%corresponding to  extra Dirac points (EDPs) between different bands. Indeed, 
then in the center  of  the Brillouin zone we have ($k_x=0, k_y = 0$), which can be injected into 
%we use 
\eqref{implicit} to get
\begin{equation}\label{EqDispl}
  \varepsilon_D \left(k_{D_x}=0,k_{D_y} = 0\right)= q_1 \mathbb{V}_1+q_2 \mathbb{V}_2+q_3 \mathbb{V}_3+2n\pi,\qquad n\in \mathbb{Z}
\end{equation}
%whereas 
and  the edge of 
 the Brillouin zone 
corresponds to $\left(k_x =\pm \frac{\pi}{d},k_y = 0\right)$, thus we obtain
%the Dirac points should appear at energy
\begin{eqnarray}\label{EqDisplpid}
  \varepsilon_D\left(k_{D_x} = \pm \frac{\pi}{d},k_{D_y} = 0\right)
 % &=& \varepsilon_D\left(k_{D_x} = -\frac{\pi}{d},k_{D_y} = 0\right)
  %\\
= q_1 \mathbb{V}_1+q_2 \mathbb{V}_2+q_3 \mathbb{V}_3+(2n\pm 1)\pi,\qquad n\in \mathbb{Z}.
\end{eqnarray}
Note that, %At this level, we have some comments in order. Indeed, 
both of \eqref{EqDispl} and \eqref{EqDisplpid} are valid for any distance $q_i\neq 1$ and potential height $\mathbb{V}_i$. Recall  that, the  pristine graphene submitted to potential $\mathbb{V}_i$ $(q_i=1)$ has only one Dirac point located at the energy  $\varepsilon_D(k_{D_x}=k_{D_y}= 0)=\mathbb{V}_i$. %Note that \eqref{EqDispl} generalizes that for SSLGSL-2R  \cite{Pham2010JPCM22}.

%For SSLGSL-3R,
The coordinates of the Dirac points  in the minibands $\left(k_x =\varepsilon = 0\right)$ can be derived from the dispersion relation as 
%to get
%This gives 
%the relation
\begin{equation}
   \hspace*{-0.5em} \frac{\mathbb{V}^2-(k_yd)^2 \cos \left((q_2-1) \sqrt{\mathbb{V}^2-
   (k_yd)^2}\right)}{\mathbb{V}^2-d^2 k_y^2}\cosh ( k_y  q_2 d)
  -\frac{ k_y d  \sin \left((q_2-1) \sqrt{\mathbb{V}^2-(k_yd)^2}\right)}
  {\sqrt{\mathbb{V}^2-(k_yd)^2}}\sinh (k_y q_2 d)=1
\end{equation}
which are strongly depending on  
%having different solutions according to the value taken by 
the distance $q_2$. Indeed, for
$q_2\neq 0$ we have %shows that there is %we have 
only one solution $k_y=0$ but for $q_2=0$ there are two %solutions
\begin{equation}\label{ADP}
  k_{D_{y}}=0, \qquad k_{D_{y}}=\pm\frac{1}{d}\sqrt{\mathbb{V}^2-(2\ell\pi)^2}
\end{equation}
and  the condition $\mathbb{V}^2>(2\ell\pi)^2$ must fulfilled with
$\ell$ is an integer no null. Then, in minibands we have extra Dirac points   ($k_x=0,k_y=k_{D_{y}},\varepsilon= 0$) in addition to the original one ($k_x=0,k_y=0,\varepsilon= 0$). % These results have been also obtained in  \cite{Pham2010JPCM22}.
 The corresponding energies can be written in compact form as
 %associated to the above solutions read as
%The Dirac points energy in the tree cases can be written as
\begin{equation}
   \varepsilon_D=0,\qquad \varepsilon_D=q_1 \mathbb{V}_1+q_2 \mathbb{V}_2+q_3 \mathbb{V}_3+m\pi,\qquad m\neq 0
   \label{eq20}.
\end{equation}
%which lead to
%At this level, we have some comments in order. 
%From \eqref{eq20}, we 
%distinguish two cases: first one is asymmetrical meaning that considering  $\varepsilon_D$ for any value of the set ($\mathbb{V}_1$, $\mathbb{V}_2$, $\mathbb{V}_3, q_1$, $q_2$, $q_3$) and second is symmetrical. In the forthcoming analysis, 

In the next, we will focus only on the case of %interested to 
symmetrical graphene with %with three regions (SSLGSL-3R) \cite{kamal2018EPJB} characterized by 
spacially modulated potential and
under the conditions $\left(\dfrac{1-q_2}{2},q_2,\dfrac{1-q_2}{2}\right)$, ($\mathbb{V}_1=-\mathbb{V}_3=\mathbb{V}$, $\mathbb{V}_2=0$). In this situation,
%which allow to write 
\eqref{eq20} reduces to the following
\begin{equation}
   \varepsilon_D=m\pi,\qquad m\in\mathbb{Z}.
\end{equation}
Now there are two cases to distinguish according to the values taken by the quantum number $m$ and the potential height $\mathbb{V}$.
First, if
$m=0$ and $\mathbb{V}\neq 0$, then for $q_2=0$ we have  the velocities
\begin{equation}\label{200}
   u_x=\left(\dfrac{2\ell\pi}{\mathbb{V}}\right)^2,\qquad u_y=1-\left(\dfrac{2\ell\pi}{\mathbb{V}}\right)^2
\end{equation}
and for $q_2\neq 0$ we get
\begin{equation}\label{201}
   u_x=1, \qquad u_y(q_2,\mathbb{V})=\frac{1}{\mathbb{V}}
  \sqrt{2+q_2^2\mathbb{V}^2-2\cos\left(\left(q_2-1\right)\mathbb{V}\right)
  -2q_2\mathbb{V}\sin\left(\left(q_2-1\right)\mathbb{V}\right)}.
\end{equation}
Second, if $m\neq 0$ and $\mathbb{V}\neq |m\pi|$, then for all $q_2$ 
we find
\begin{eqnarray}\label{202}
   u_x&=&1\\ 
   u_y^\pm&=&\frac{\mathbb{V}}{\sqrt{2}m\pi\left|(m\pi)^2-\mathbb{V}^2\right|}\Big[3(m\pi)^2+
  \mathbb{V}^2+((m\pi)^2-\mathbb{V}^2)\cos(2m\pi q_2)\nonumber\\
  &&\pm2m\pi\Big(\left(m\pi+\mathbb{V}\right)\cos\left(m\pi q_2+\left(1-q_2\right)\mathbb{V}\right)+
  \left(m\pi-\mathbb{V}\right)\cos\left(m\pi q_2-\left(1-q_2\right)\mathbb{V}\right)\Big)\Big]^{1/2}\nonumber 
\end{eqnarray}
where $\pm$ refer to  odd and even integers. To allow for a better understanding, 
we present the obtained results 
 %we summarize
%the coordinates of Dirac points (${k_D}_x$, ${k_D}_y$, $\varepsilon_D$) and  corresponding coefficients ($u_x$, $u_y$) for  $\mathbb{V}\neq m\pi$ 
in Table \ref{Tab:fxfy}:\\

\begin{table}[!htb]
\centering
\renewcommand{\arraystretch}{2}
\begin{tabular}{cc|c|c|c|c|c|}
\cline{3-7}
& & \multicolumn{3}{c}{Extra Dirac points (${k_D}_x$, ${k_D}_y$, $\varepsilon_D$)}& \multicolumn{2}{|c|}{Velocity components ($u_x$,$u_y$)}
\\
\cline{3-7}
&  & ${k_D}_x$  & ${k_D}_y$  & $\varepsilon_D$ & $u_x$ & $u_y$ \\
\hline
\multicolumn{1}{|c|}{\multirow{3}{*}{$m=0$}} & \multirow{2}{*}{$q_2=0$} & $0$ & $\pm\dfrac{\sqrt{\mathbb{V}^2-(2\ell\pi)^2}}{d}$, $\ell\neq 0$ & $0$ & $\left(\dfrac{2\ell\pi}{\mathbb{V}}\right)^2$ & $1-\left(\dfrac{2\ell\pi}{\mathbb{V}}\right)^2$ \\
\cline{3-7}
\multicolumn{1}{|c|}{} &  & \multirow{1}{*}{$0$}  & \multirow{1}{*}{$0$} & \multirow{1}{*}{$0$}& \multirow{1}{*}{$1$} & \multirow{1}{*}{$u_y(0,\mathbb{V})$} \\
\cline{2-7}
\multicolumn{1}{|c|}{} & \multicolumn{1}{|c|}{$q_2\neq0$} & $0$ & $0$ & $0$ & $1$ & $u_y(q_2,\mathbb{V})$ \\
\hline
\multicolumn{1}{|c|}{\multirow{2}{*}{$m\neq0$}} & \multicolumn{1}{|c|}{$m=2n$, $n\neq0$} & $0$ & $0$ & $m\pi$ & \multirow{2}{*}{1}                                                 & $u_y^-$ \\
\cline{2-5}\cline{7-7}
\multicolumn{1}{|c|}{} & \multicolumn{1}{|c|}{$m=2n\pm1$} & \multicolumn{1}{|c|}{$\pm\dfrac{\pi}{d}$} & \multicolumn{1}{c|}{$0$} & \multicolumn{1}{|c|}{$m\pi$} &  & $u_y^+$ \\ \hline
\end{tabular}
\caption{Coordinates of the extra Dirac points  (${k_D}_x$, ${k_D}_y$, $\varepsilon_D$) and the corresponding velocity components ($u_x$, $u_y$) for  Dirac fermions in graphene with spacially modulated potential.}
\label{Tab:fxfy}
\end{table}

%============================================
\section{Zitterbewegung effect}\label{ZB}
%============================================

We will show that
in the first Brillouin zone, graphene with spacially modulated potential could influence the ZB effect under the change of the  group velocities.
For this, we  
study the dynamic of fermions 
%, which
% are 
described by the Hamiltonian \eqref{eq:M_prime} near EDPs together with ODP. Indeed,  we use the Heisenberg formalism to introduce the time evolution of  the position operators $x(t)$ and $y(t)$ %evolving with time $t$ are given by the transformations
\begin{equation}
x(t)= e^{-iHt/\hbar}x(0)e^{iHt/\hbar},\qquad y(t)= e^{-iHt/\hbar}y(0)e^{iHt/\hbar}
\label{eq:rt}
\end{equation}
satisfying the Heisenberg equations of motion
\begin{equation}
 \dfrac{dx}{dt}=%\dfrac{i}{\hbar}\left[H,x\right]=
 u_x v_F \sigma_x,\qquad \dfrac{dy}{dt}=%\dfrac{i}{\hbar}\left[H,y\right]=
 u_y v_F \sigma_y.
\label{eq:dr}
\end{equation}
To explicitly determine
%To get the explicit form of 
$x(t)$ and $y(t)$, we consider the dynamics of Pauli operators $(\sigma_x,\sigma_y)$
and show the relations
\begin{equation}\label{eq27}
 -i\hbar\dfrac{d\sigma_x}{dt}= 2H\eta_x,
 %\left[H,\sigma_x\right],
 \qquad  -i\hbar\dfrac{d\sigma_y}{dt}= %\left[H,\sigma_y\right]
 2H\eta_y
\end{equation}
%and 
%using the anti-commutation relations $\left\{\sigma_i,\sigma_j\right\}=2\delta_{ij}\mathbb{I}$ $(i,j=x,y,z)$, we write %\eqref{eq27} as% and obtain the compact relations
%\begin{equation}
 %-i\hbar\dfrac{d\sigma_x}{dt}=2H\eta_x,\qquad -i\hbar\dfrac{d\sigma_y}{dt}=2H\eta_y
 %\label{eq:dsigma}
%\end{equation}
such that
%and we have 
the operators $\eta_x$ and $\eta_y$
are given by
\begin{equation}
 \eta_x=\sigma_x-\hbar u_x v_F \kappa_xH^{-1}-\varepsilon_DE_FH^{-1}\sigma_x,\qquad \eta_y=\sigma_y-\hbar u_y v_F \kappa_yH^{-1}-\varepsilon_DE_FH^{-1}\sigma_y
 \label{eq:eta}
\end{equation}
where the wave vector $\bm{\kappa}=\bm{k}-\bm{k_D}$ has two components
($\kappa_x=k_x-k_{D_{x}}$, $\kappa_y=k_y-k_{D_{y}}$). One can show that the  dynamics
equations of $\eta_x$ and $\eta_y$ take the forms
\begin{equation}
 -i\hbar\dfrac{d\eta_x}{dt}=2H\eta_x,\qquad -i\hbar\dfrac{d\eta_y}{dt}=2H\eta_y
 \label{eq:deta}
\end{equation}
which can be solved to end up with %the solutions
%whose solutions are given by
\begin{equation}
 \eta_x(t)=e^{2i(H-\varepsilon_DE_F\mathbb{I})t/\hbar}{\eta_0}_x,\quad \eta_y(t)=e^{2i(H-\varepsilon_DE_F\mathbb{I})t/\hbar}{\eta_0}_y.
 \label{eq:etat2}
\end{equation}
%Combining 
Using this together with
%Using 
\eqref{eq:eta} %and \eqref{eq:etat2}, we 
to write the dynamics \eqref{eq:dr} as
%to end up with
\begin{eqnarray}
 \dfrac{dx}{dt}&=&\hbar v_F^2 u_x^2\kappa_x (H-\varepsilon_DE_F\mathbb{I})^{-1}+u_x v_F H(H-\varepsilon_DE_F\mathbb{I})^{-1}e^{2i(H-\varepsilon_DE_F\mathbb{I})t/\hbar}{\eta_0}_x\label{eq:dr21}\\
 \dfrac{dy}{dt}&=&\hbar v_F^2 u_y^2\kappa_y (H-\varepsilon_DE_F\mathbb{I})^{-1}+u_y v_F H(H-\varepsilon_DE_F\mathbb{I})^{-1}e^{2i(H-\varepsilon_DE_F\mathbb{I})t/\hbar}{\eta_0}_y
\label{eq:dr22}
\end{eqnarray}
%which can be integrated to get 
%\begin{eqnarray}
 %&& x(t)=x(0)+\hbar v_F^2 \alpha^2\kappa_x (H-\varepsilon_DE_F\mathbb{I})^{-1}t+\dfrac{i\hbar}{2}\alpha v_F H(H-\varepsilon_DE_F\mathbb{I})^{-2}\left(\mathbb{I}-e^{2i(H-\varepsilon_DE_F\mathbb{I})t/\hbar}\right){\eta_0}_x
%\label{eq:rt21}
%\\
%&& y(t)=y(0)+\hbar v_F^2 \beta^2\kappa_y (H-\varepsilon_DE_F\mathbb{I})^{-1}t+\dfrac{i\hbar}{2}\beta v_F H(H-\varepsilon_DE_F\mathbb{I})^{-2}\left(\mathbb{I}-e^{2i(H-\varepsilon_DE_F\mathbb{I})t/\hbar}\right){\eta_0}_y
%\label{eq:rt22}
%\end{eqnarray}
%where we have set
%with the quantities
%\begin{eqnarray}
 %{\eta_0}_x&=&\sigma_x(0)-\hbar \alpha v_F \kappa_xH^{-1}-\varepsilon_DE_FH^{-1}\sigma_x(0)\label{eq:eta021}\\
 %{\eta_0}_y&=&\sigma_y(0)-\hbar \beta v_F \kappa_yH^{-1}-\varepsilon_DE_FH^{-1}\sigma_y(0)
 %\label{eq:eta022}
%\end{eqnarray}
%$x(0)$ and $y(0)$ are constant operators of integration. Substituting \eqref{eq:eta021}-\eqref{eq:eta022} into \eqref{eq:rt21}-\eqref{eq:rt22}, we obtain the analytical expression
%After integration,  we find the solutions in matrix representation
and their solutions can be worked out to find
\begin{eqnarray}
 x(t) &=&x_0\mathbb{I}+u_x v_Ft\sigma_x+\frac{u_xu_y\kappa_y}{2\kappa^2_e}\left[1-\cos{(2v_F\kappa_e t)}\right]\sigma_z\\
 && +\dfrac{u_xu_y \kappa_y}{2k_e^3}\left[2v_F\kappa_e t-\sin{\left(2v_F\kappa_e t\right)}\right]\left(u_x\kappa_x \sigma_y-u_y\kappa_y \sigma_x\right) \nonumber\\
y(t) &=& y_0\mathbb{I}-u_x v_Ft\sigma_y+\frac{u_xu_y\kappa_y}{2\kappa^2_e}\left[1-\cos{(2v_F\kappa_e t)}\right]\sigma_z  \\
&& +\dfrac{u_xu_y \kappa_y}{2k_e^3}\left[2v_F\kappa_e t-\sin{\left(2v_F\kappa_e t\right)}\right]\left(u_x\kappa_x \sigma_y-u_y\kappa_y \sigma_x\right)
 \label{eq:r} \nonumber
\end{eqnarray}
%It is clearly seen that 
where 
$x_0$ and $y_0$ are  the constant operators of integration. It is clearly
seen that
the two first matrix elements %of matrix representation corresponding to the position operators
%take the forms
are given by
\begin{eqnarray}
 {x}_{11}(t)=x_0+\frac{u_xu_y\kappa_y}{2\kappa_e^2}[1-\cos{(2v_F\kappa_e t)}]
 \label{eq411}
\\
 y_{11}(t)=y_0-\frac{u_xu_y\kappa_x}{2\kappa_e^2}[1-\cos{(2v_F\kappa_e t)}]
 \label{eq42}
\end{eqnarray}

Now we proceed by evaluating the averages of time-dependent position operators within a Gaussian wave packet %by considering %the condition that
by assuming that the initial state of Dirac fermions is described by 
the spinor %a two-dimensional wave packet
%\cite{Schliemann2006PRB} 
\cite{Schliemann2005PRL}
\begin{equation}\label{stee}
 \psi(\bm{r},0)=\frac{1}{2\pi}\frac{\sigma}{\sqrt{\pi}}\int d^2\bm{\kappa}~e^{-\frac{1}{2}\sigma^2(\bm{\kappa}-\bm{\kappa}_0)^2}e^{i\bm{\kappa}\cdot \bm{r}}
 \left(\begin{array}{c}1\\0\end{array} \right)
\end{equation}
where the unit vector is a convenient choice, 
$\sigma$ 
 and 
 $\bm{\kappa_0}=(\mathit{\kappa_{0x}},\mathit{\kappa_{0y}})$
 are the width and center of the wave packet, respectively. 
 %The unit vector $\begin{pmatrix}1\\0\end{pmatrix}$ is a convenient choice \cite{Schliemann2005PRL} and 
 Then, the averages of the two matrix elements with respect to the spinor \eqref{stee} 
 are found to be of the forms 
 %${x}_{11}(t)$ and ${y}_{11}(t)$ are given by %take the forms
\begin{eqnarray}
 &&\bar{x}_{11}(t) %&=&\langle{\psi}|x(t)|{\psi}\rangle\nonumber\\
 =\frac{\sigma^2}{\pi}\int d^2\bm{\kappa}\  \frac{u_xu_y\kappa_y}{2\kappa_e^2}\left[1-\cos{(2v_F\kappa_e t)}\right]e^{-\sigma^2(\bm{\kappa}-\bm{\kappa}_0)^2}
 \label{eq:x111}
\\
 &&\bar{y}_{11}(t) %&=&\langle{\psi}|y(t)|{\psi}\rangle\nonumber\\
 =-\frac{\sigma^2}{\pi}\int d^2\bm{\kappa}\ \frac{u_xu_y\kappa_x}{2\kappa_e^2}\left[1-\cos{(2v_F\kappa_e t)}\right]e^{-\sigma^2(\bm{\kappa}-\bm{\kappa}_0)^2}
 \label{eq:x112}.
\end{eqnarray}
It is clearly seen that both of expressions are 
involving 
the ZB frequency $\omega(\kappa)=2v_F\kappa_e$ and explicitly we have
\begin{equation}
 \omega(\kappa)=2\kappa v_F \sqrt{u_x^2\cos^2\theta_0+u_y^2\sin^2\theta_0}
 \label{freq}
\end{equation}
where the angle $\theta$ is given in terms of the wave vector
\begin{equation}
e^{i\theta_0}= \frac{\kappa_x + i\kappa_y}{\kappa} = 
\frac{u_x\kappa_x + iu_y\kappa_y}{\kappa_e}
\end{equation}
such that $\kappa_e= (u_x\kappa_x, u_y\kappa_y)$.
Note  that, \eqref{freq} can be
determined by the difference between the upper and lower energy branches for a given wave vector  $\bm{\kappa}$. It is convenient 
for the numerical uses to consider  %introduce 
the average
%In the following, we will focus on the ZB effect in the vicinity of  Dirac points by analyzing the oscillations of the modulo of 
%average 
of the position operator $\bar{r}(t)$ %defined by
\begin{equation}
 \bar{r}(t)=\bar{x}_{11}(t)+\mathrm{i}~\bar{y}_{11}(t).
 \label{posRt}
\end{equation}
According to expressions (\ref{200}-\ref{202}) taken by
%Since 
the velocities $u_x$ and $u_y$, we will see that
%are functions of the distance $q_2$ and potential $\mathbb{V}$ %(except in the case where $m=0$, $q_2=0$ they are function of $\ell$ and $\mathbb{V}$). The relation 
%then
\eqref{freq} will provide a convenient way for adjusting the ZB oscillations 
by tuning on 
%via the applied 
the potential height
$\mathbb{V}$ and distance $q_2$ of the central region. 
%and initial momentum of Gaussian wave packet ($\kappa_0$, $\theta_0$).

%In the following analysis, we will study the oscillations of its modulo under suitable conditions
%of the physical parameters

%===============================
\section{Numerical results}\label{Results}
%===============================

To analyze the influence of spacially modulated potential on the ZB effect,
we numerically analyze the averages of position operators $\bar x_{11}$ \eqref{eq:x111},
$\bar y_{11}$ \eqref{eq:x112} and  $|\bar r|$ \eqref{posRt}
near original Dirac point (ODP) and
extra Dirac points (EDPs)   under suitable conditions of the set of parameters ($q_2$, $\mathbb{V}$, $d$, $m$, $\sigma$, $\kappa_0$, $\theta_0$) characterizing the applied potential and Gaussian wave packet. To carry out our computations, we take an appropriate %wave packet 
width $\sigma$ that allows investigating the influence of each extra Dirac point on the ZB
effect. To avoid the ZB effect being influenced by all Dirac points and make 
%wave packet width 
$\sigma$ not too large in momentum space, we choose one cell distance such that $d=200~a$ as function of the interatomic distance $a=0.142$ nm. 
%The numerical results will be analyzed and presented in 11 Figures \ref{FigrODPq20Theta}-\ref{FigrODPq20Theta}

\begin{figure}[!hbt]\centering
    \subfloat[]{\centering
    \includegraphics[width=8cm]{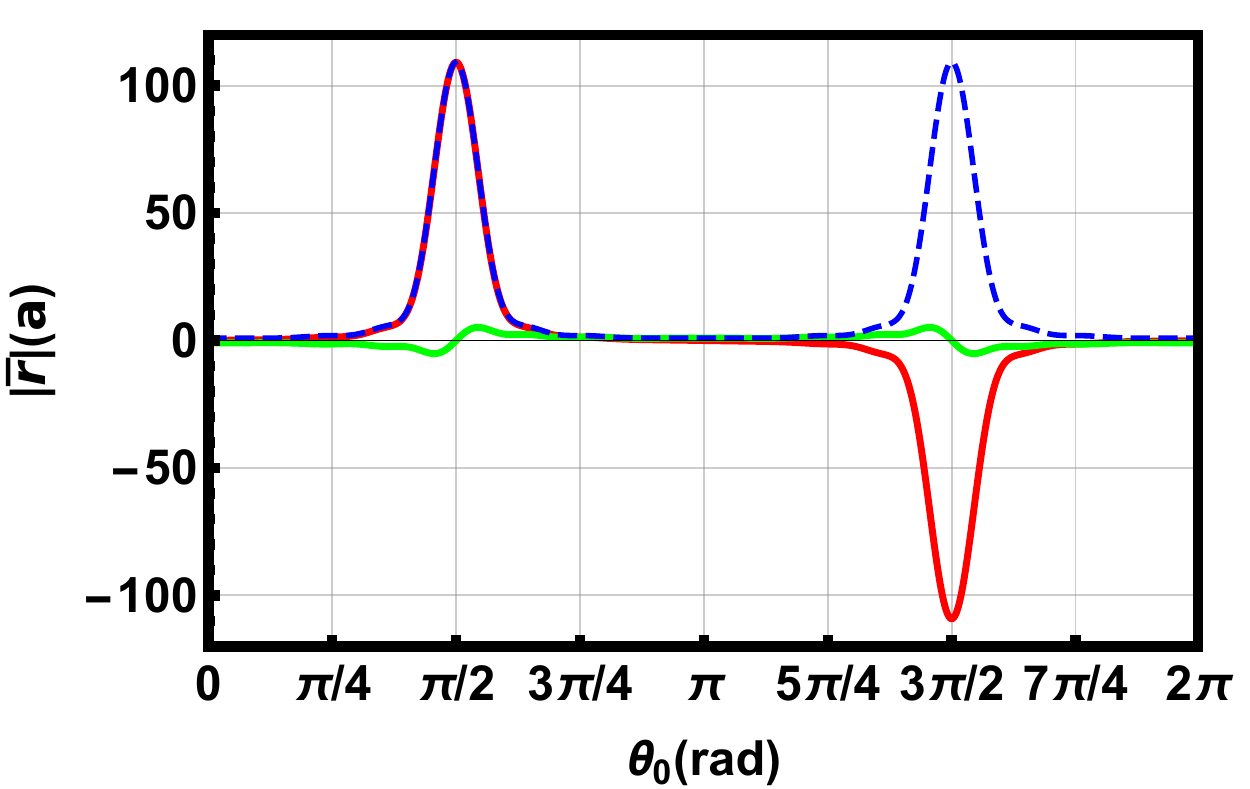}
    \label{Fig3a}
    }
    \subfloat[]{\centering
    \includegraphics[width=8cm]{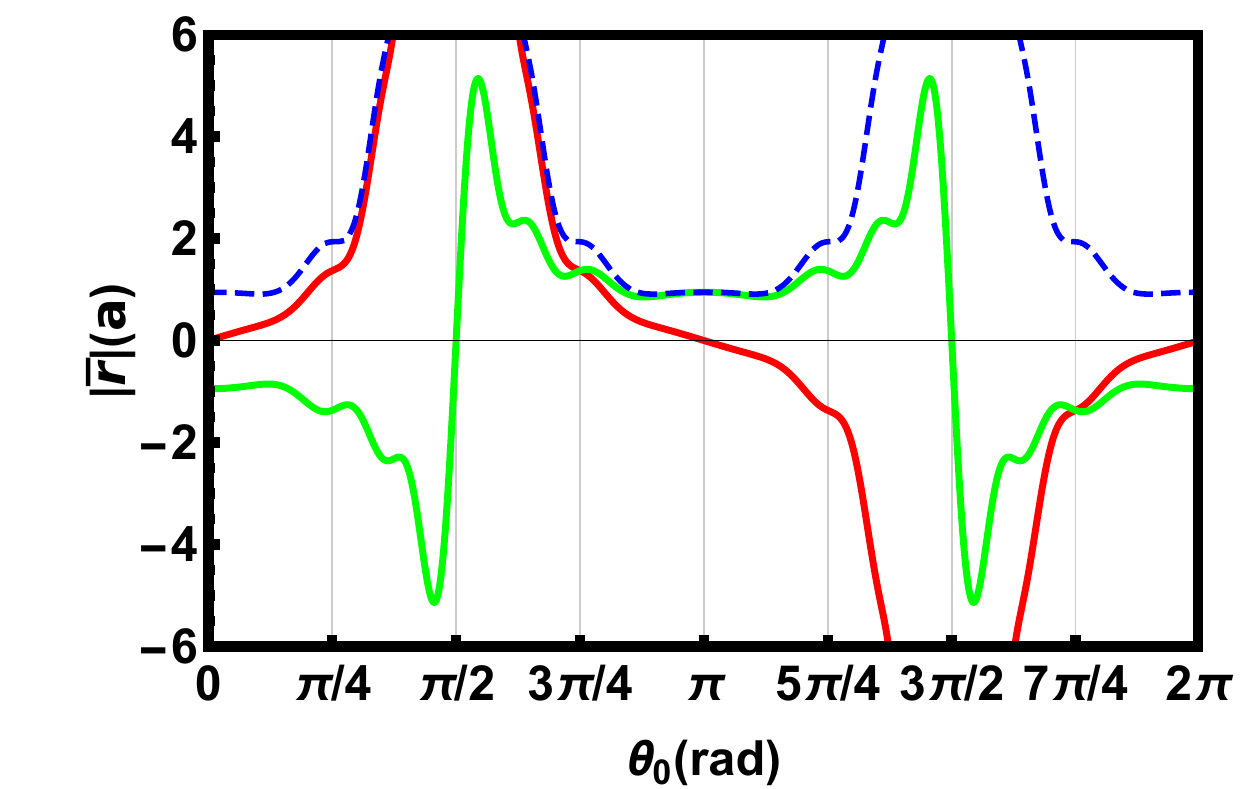}
    \label{Fig3b}
    }
    \caption{
    (color online) \protect\subref{Fig3a}: The averages of position operators 
$\bar x_{11}$ (red), $\bar y_{11}$ (green) and $|\bar r|$ (blue) 
    %The ZB oscillation 
    versus the angle $\theta_0$ near ODP for $t=50~\femto\second$, $\mathbb{V}=7\pi$, $q_2=0$, $\kappa_0=0.05~a^{-1}$, $\sigma=150~a$.  \protect\subref{Fig3b}: The behaviors of three averages  for small values of the positions $[-6a,6a]$.
    %presented in \protect\subref{Fig3a} when the position belongs to $[-6a,6a]$.
    %The $x$-direction  (green line), $y$-direction  (red line) and  $\bar{r}$ (blue dashed) ZB oscillations.
    }
    \label{FigrODPq20Theta}
\end{figure}

Figure \ref{FigrODPq20Theta} presents 
%In Figure \ref{FigrODPq20Theta}\protect\subref{Fig3a}, we present 
%In the following, we analyze 
the ZB oscillations of the averages of position operators
$\bar x_{11}$,
$\bar y_{11}$  and  $|\bar r|$ 
%of the displacement of an electron $|\bar{r}(t)|$
versus the angle 
$\theta_0$ for  time $t=50~\femto\second$ 
near 
ODP, i.e. ${k_D}_x={k_D}_y=\varepsilon_D=0$. 
%We present in Figure \ref{FigrODPq20Theta} the ZB oscillations 
%of the displacement 
%as function of the angle $\theta_0$ for  time $t=50~\femto\second$.
In Figure \ref{FigrODPq20Theta}\protect\subref{Fig3a}, 
we observe  that the $x$-direction ZB oscillation is predominate compared to the $y$-direction one
%ZB oscillation. 
with an amplitude %of $y$-direction ZB oscillation is 
very large in the vicinity of $\theta_0=\pi/2~[\pi]$. 
%It is clearly seen % addition, we observe that
In addition, there are two 
%that there are some
symmetries with respect to $\theta_0=\pi$ such that
$|\bar r|$ is showing symmetrical behavior while $\bar x_{11}$ and $\bar y_{11}$
are presenting asymmetrical ones.
To shed light on the averages behaviors for small positions, 
%
%For clarification, 
%To better understand such behavior 
we zoom one part in Figure \ref{FigrODPq20Theta}\protect\subref{Fig3b} to show that 
%showing that 
 $\bar y_{11}$ presents also some pics.

\begin{figure}[!hbt]\centering
   \subfloat[]{\centering
      \includegraphics[width=5.59cm]{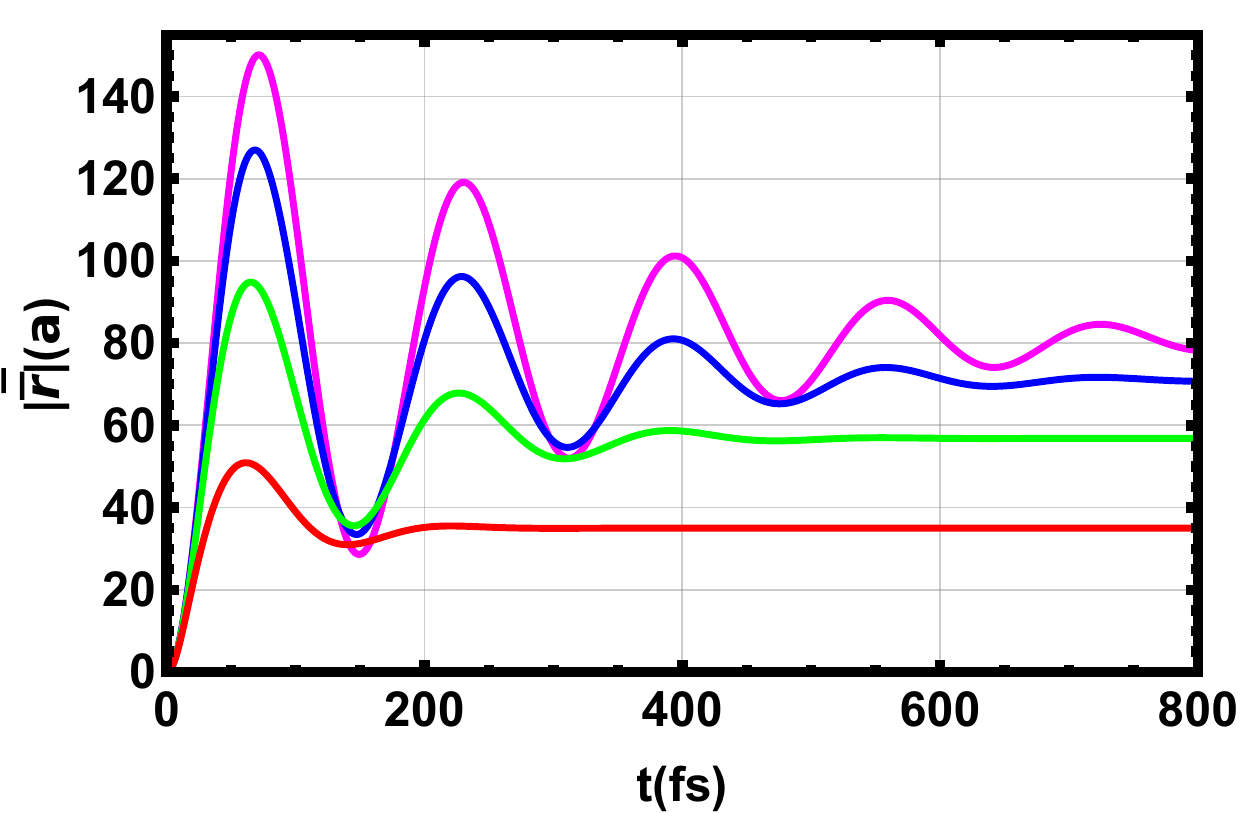}
      \label{FigrODPq20:SubFigA}
   }
   \subfloat[]{\centering
      \includegraphics[width=5.59cm]{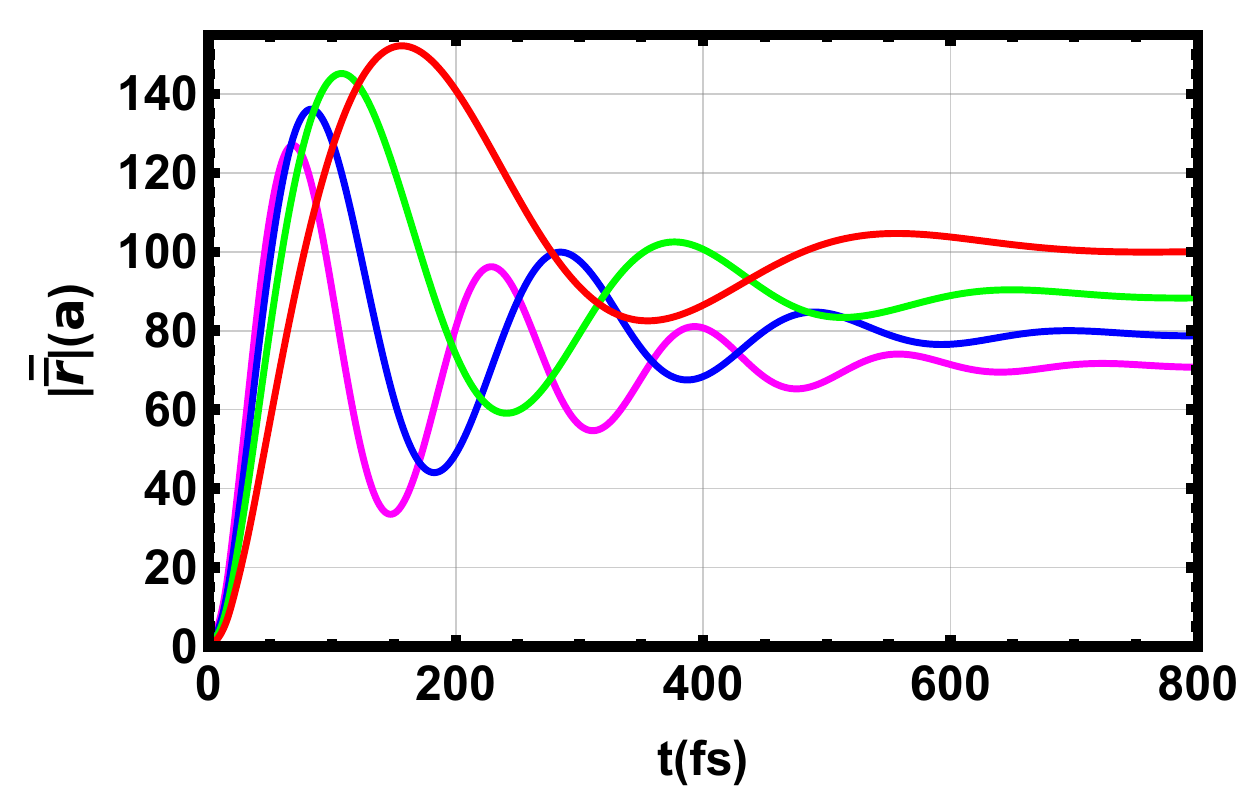}
      \label{FigrODPq20:SubFigB}
   }
   \subfloat[]{\centering
      \includegraphics[width=5.59cm]{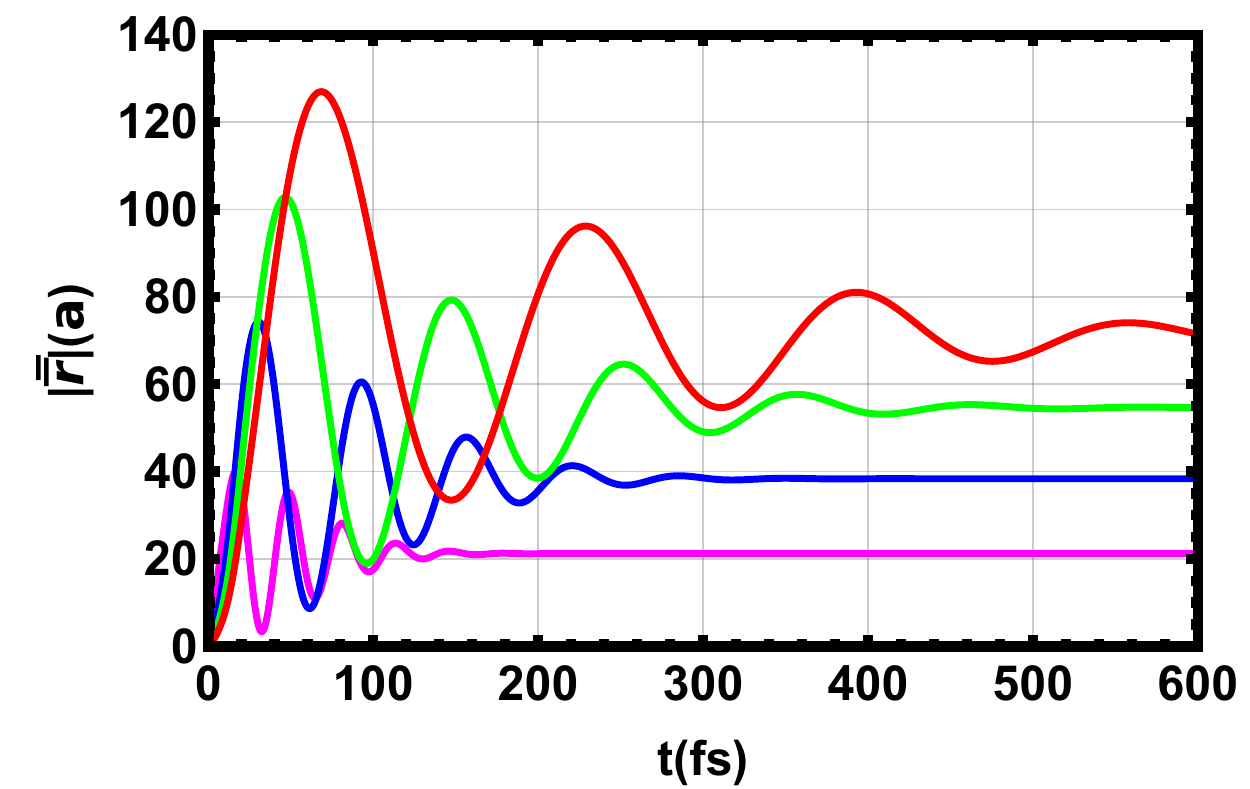}
      \label{FigrODPtq2:SubFigC}
   }
   \caption{
   (color online) The average of position operator $|\bar r|$ versus the time near ODP with $\mathbb{V}=7\pi$, $\theta_0=\pi/2$. \protect\subref{FigrODPq20:SubFigA}: $q_2=0$, $\kappa_0=0.05~a^{-1}$ for different values of the packet width %of Gaussian wave packet
   $\sigma= 50~a^{-1}$ (red), $100~a^{-1}$ (green), $150~a^{-1}$ (blue), $200~a^{-1}$ (magenta). \protect\subref{FigrODPq20:SubFigB}: $q_2=0$, $\sigma= 150~a^{-1}$ for different values of the packet center  $\kappa_0= 0.02~a^{-1}$ (red), $0.03~a^{-1}$ (green), $0.04~a^{-1}$ (blue), $0.05~a^{-1}$ (magenta). \protect\subref{FigrODPtq2:SubFigC}: $\sigma= 150~a^{-1}$, $\kappa_0=0.05~a^{-1}$ for different values of the distance $q_2= 0$ (red), $1/8$ (green), $1/4$ (blue), $1/2$ (magenta).
   }
    \label{FigrODPq20}
\end{figure}

%\newpage

%To have a very good manifestation of the  ZB effect, we study the ZB oscillations for $\theta_0=\pi/2$. 
Figure \ref{FigrODPq20} %\subref{FigrODPq20:SubFigA} 
shows the ZB oscillations of 
the average of position operator $|\bar r|$
%displacement of electrons
as function of time near ODP for %a given Gaussian wave packet with $\kappa_0=0.05~a^{-1}$ 
$\mathbb{V}=7\pi$, $\theta_0=\pi/2$ and different values of the parameters $(\sigma,\kappa_0, q_2)$. Indeed, in Figure \ref{FigrODPq20}\subref{FigrODPq20:SubFigA} 
with $\kappa_0=0.05~a^{-1}$ and $q_2=0$,
we clearly see that by increasing  the packet width $\sigma$, the amplitude of the ZB oscillations becomes large and the attenuation becomes slow. In fact, for small $\sigma$ there are no oscillations while for very large $\sigma$ the ZB oscillations are nearly undamped. We notice that the period of the ZB oscillations is weakly depending on $\sigma$, 
%which is not the case 
%for but 
but the corresponding amplitudes are strongly depending  on $\sigma$, which is 
%This result is 
consistent with the previous analysis in literature \cite{Zhang2008PRL,Wang2014PRA}. In Figure \ref{FigrODPq20}\subref{FigrODPq20:SubFigB} with $\sigma=150~a^{-1}$ and $q_2=0$, 
%we present ZB oscillations of the displacement as a function of time near the ODP for different values of $\kappa_0$ with $q_2=0$, $\sigma=150~a$, and $\theta_0=0$. 
it is found that the amplitude and period of the ZB oscillations decrease, but the attenuation remains constant. This can be explained by the fact that the initial normalized frequency $\omega=\omega(\kappa_0)$ of the ZB oscillations 
%of the electrons displacement 
takes the form
\begin{equation}\label{fuy}
\dfrac{\omega}{\omega_0}=\dfrac{\sqrt{2-2\cos\mathbb{V}}}{\mathbb{V}}, \qquad \omega_0=2\kappa_0 v_F
\end{equation}
which  exactly coincides %in fact
%which is 
with the normalized group velocity %$v_y/v_F$ 
$u_y$ along the $y$-direction that can be obtained from \eqref{201} 
by requiring $q_2=0$.
This clearly shows that  
%has been found in \cite{kamal2018EPJB} shows that 
the velocity $u_y$ affects the ZB oscillations near ODP. To observe the ZB effect in the vicinity of ODP, we should not apply a potential of the form $\mathbb{V}=2n\pi$ ($n$ is integer) because according to \eqref{fuy} the %ZB oscillation 
frequency $\omega$ will be null. Contrariwise, for such potential $\mathbb{V}=2n\pi$ and from \eqref{freq} we show that the frequency is of order $10^{13}~\hertz$. In Figure \ref{FigrODPq20}\subref{FigrODPtq2:SubFigC} with $\sigma= 150~a^{-1}$ and $\kappa_0=0.05~a^{-1}$,
%illustrates ZB oscillations of the displacement as function of time with $\sigma= 150~a^{-1}$ for different $q_2$. As 
we observe that as the distance increases from $q_2 = 0$ to $q_2 =1/2$, the period,  amplitude and  attenuation of the ZB oscillations outstandingly reduce.

\begin{figure}[!htb]\centering
    \subfloat[]{\centering
         \includegraphics[width=7.5cm]{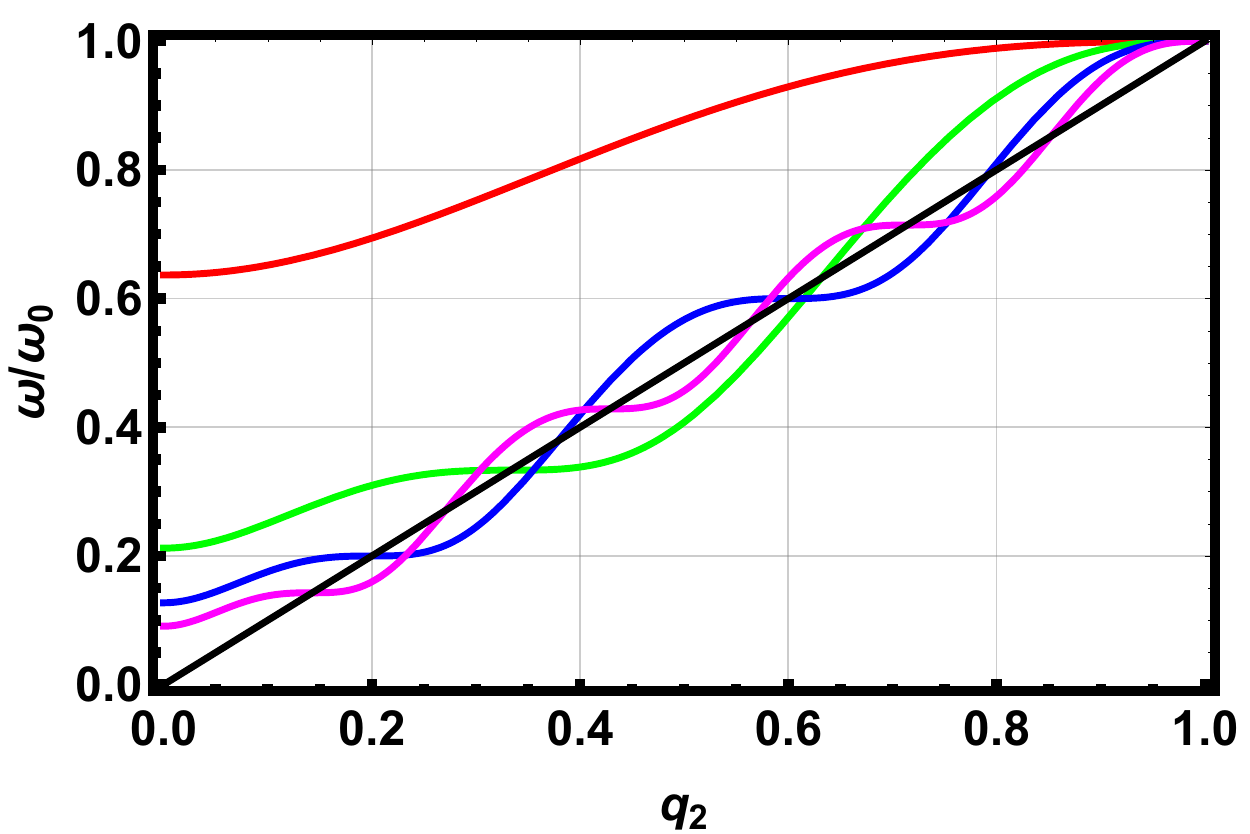}
 		\label{FigODP_Omegaq2:SubFigA}
 	}
 	\subfloat[]{\centering
         \includegraphics[width=7.5cm]{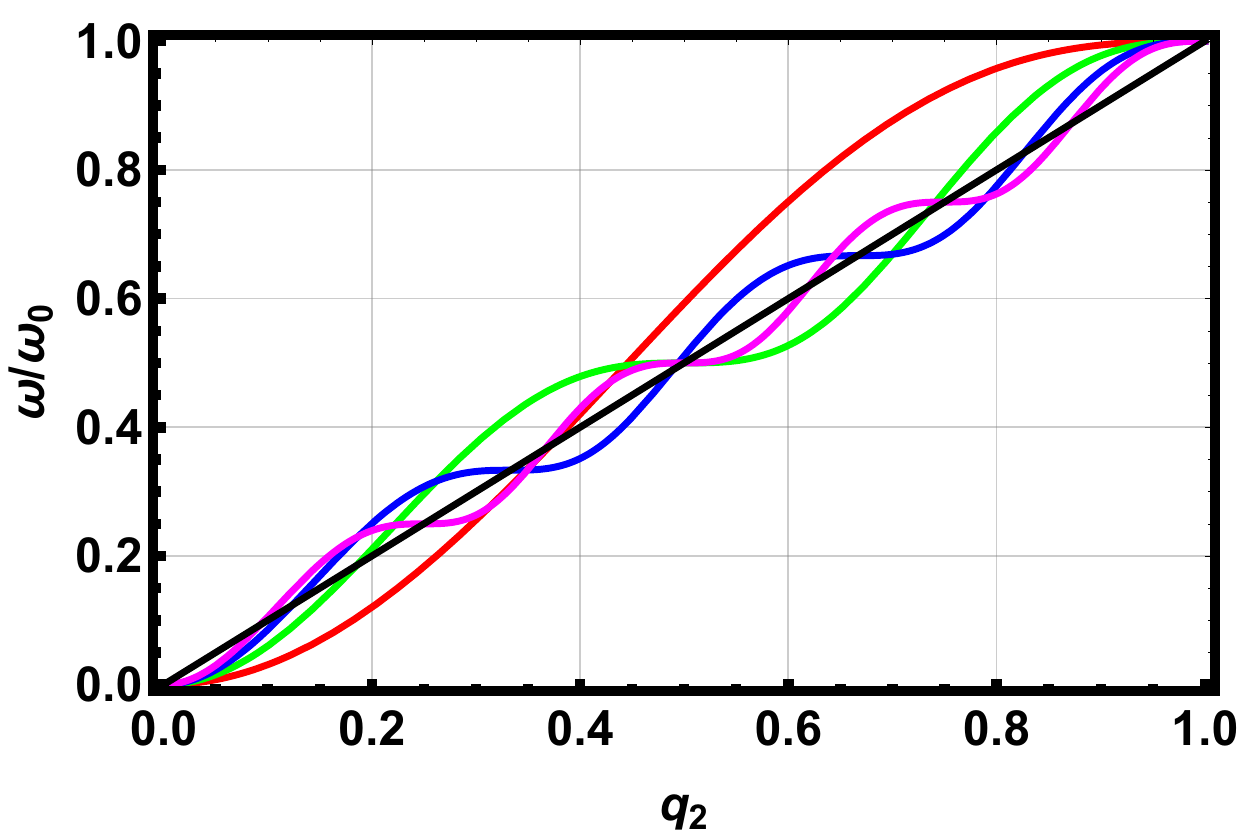}
 		\label{FigODP_Omegaq2:SubFigB}
 	}
 \caption{
    (color online) The initial normalized  frequency $\omega/\omega_0$ of the ZB oscillations versus the distance $q_2$ for $\mathbb{V}=m\pi$, with $m$ is integer. \protect\subref{FigODP_Omegaq2:SubFigA}: Odd values  $m=1$ (red), $3$ (green), $5$ (blue), $7$ (magenta). \protect\subref{FigODP_Omegaq2:SubFigB}: Even values $m=2$ (red), $4$ (green), $6$ (blue), $8$ (magenta). In both Figures the limiting case $\mathbb{V}\longrightarrow \infty$ (black) is considered.
 }
 \label{ODP_Omegaq2}
\end{figure}

Now we examine the case where
%By fixing 
the angle is fixed as $\theta_0=\pi/2$. Then,  
the initial normalized frequency of the ZB oscillations \eqref{freq} reduces to
\begin{equation}
 \dfrac{\omega}{\omega_0}=\frac{1}{\mathbb{V}}
  \sqrt{2+q_2^2\mathbb{V}^2-2\cos\left(\left(q_2-1\right)\mathbb{V}\right)
  -2q_2\mathbb{V}\sin\left(\left(q_2-1\right)\mathbb{V}\right)}
\end{equation}
which is 
%This will be 
plotted in Figure \ref{ODP_Omegaq2} 
%the initial normalized  frequency $\omega/\omega_0$ of the ZB oscillations 
versus the distance $q_2$ for $\mathbb{V}=7\pi$. For $m$ an odd integer and $q_2=0$, the frequency is nonzero and decreases as long as $\mathbb{V}$ increases. However, for any  even integer value of $m$, the frequency is zero for $q_2=0$. On the other hand, it is clearly seen that when $q_2$ increases $\omega/\omega_0$ also increases. In addition, the frequency $\omega/\omega_0$ oscillates around the straight line $\omega/\omega_0=q_2$, but for a large value of the potential height $\mathbb{V}$, we have exactly the convergence $\omega/\omega_0=q_2$. These results tell us that
%$\omega$ goes to $q_2\omega_0$, 
%which shows that 
the ZB oscillations in the vicinity of ODP can be controlled by 
the parameters $\mathbb{V}$ and $q_2$.

\begin{figure}[ht]\centering
    \subfloat[]{
        \includegraphics[width=5.58cm]{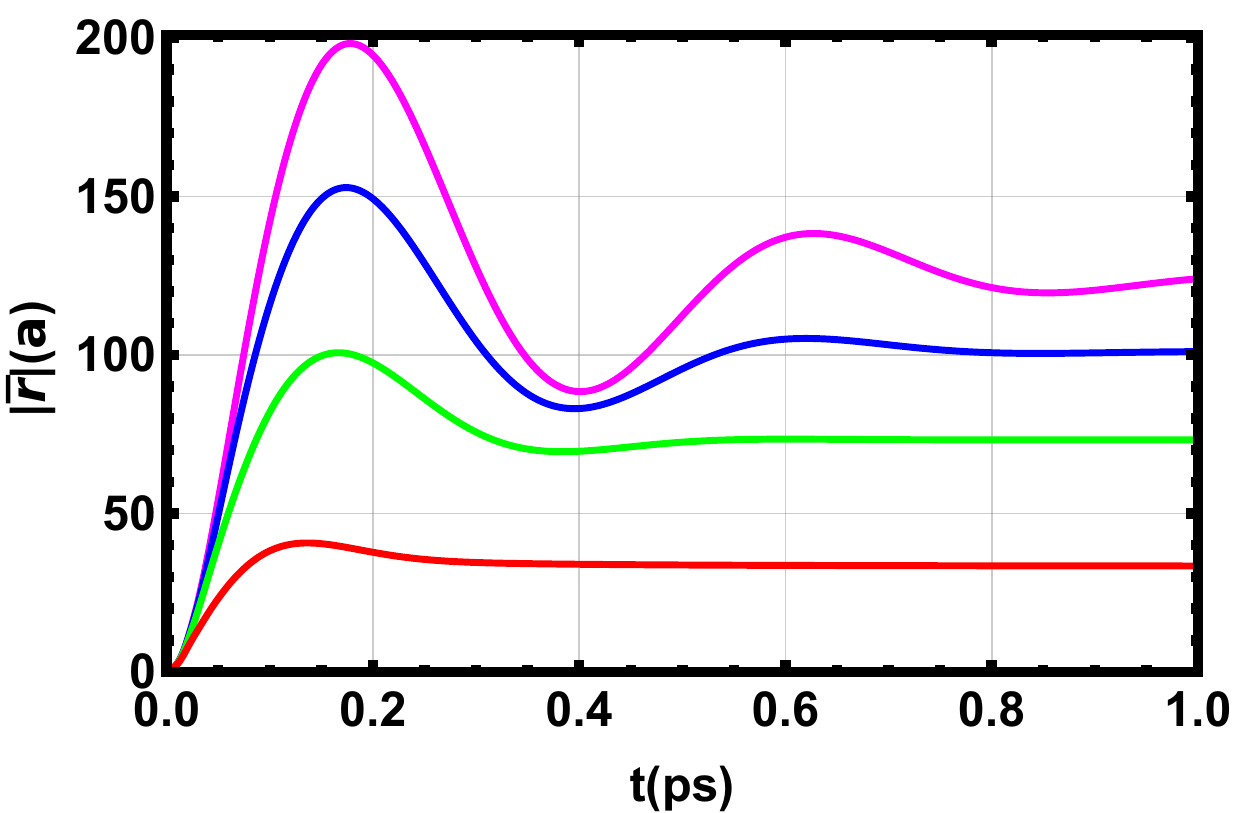}
        \label{FigrADPq20:SubFigA}
    }
    \subfloat[]{
        \includegraphics[width=5.58cm]{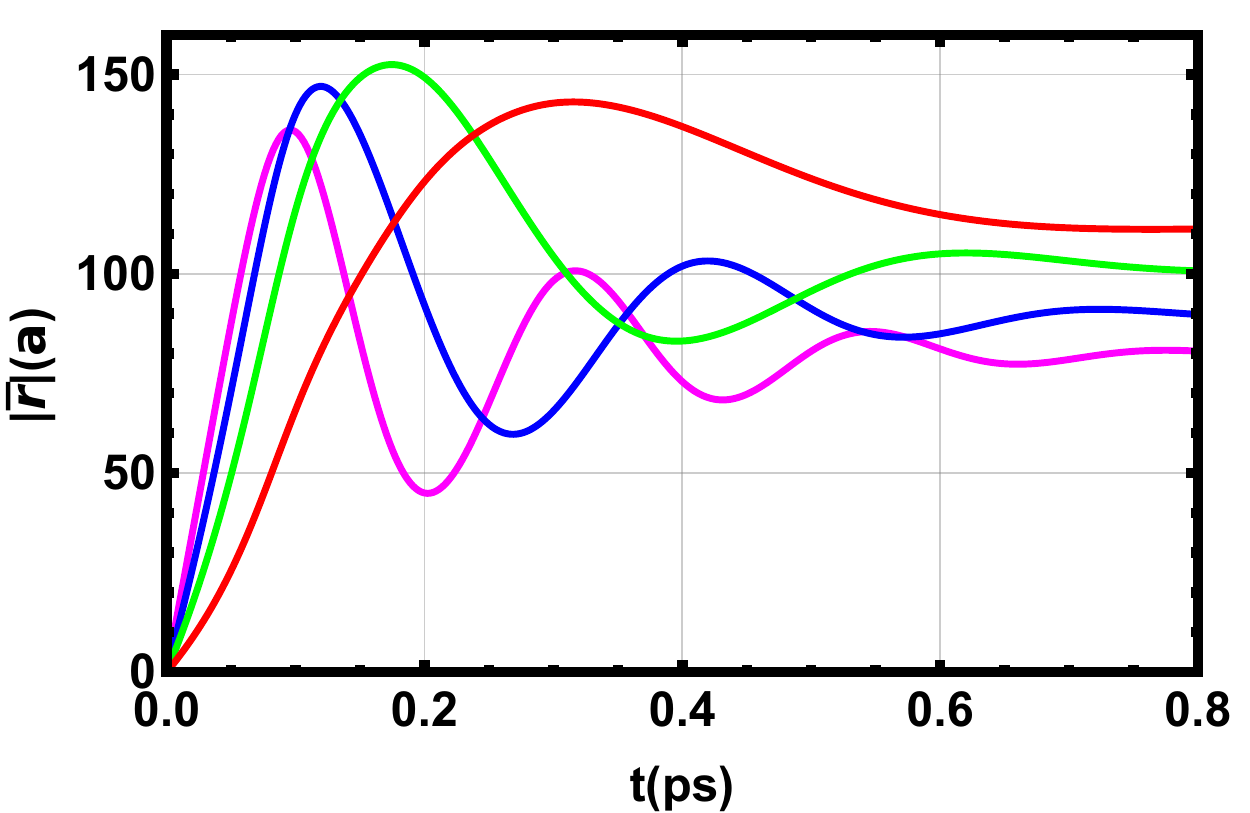}
        \label{FigrADPq20:SubFigB}
    }
    \subfloat[]{
        \includegraphics[width=5.58cm]{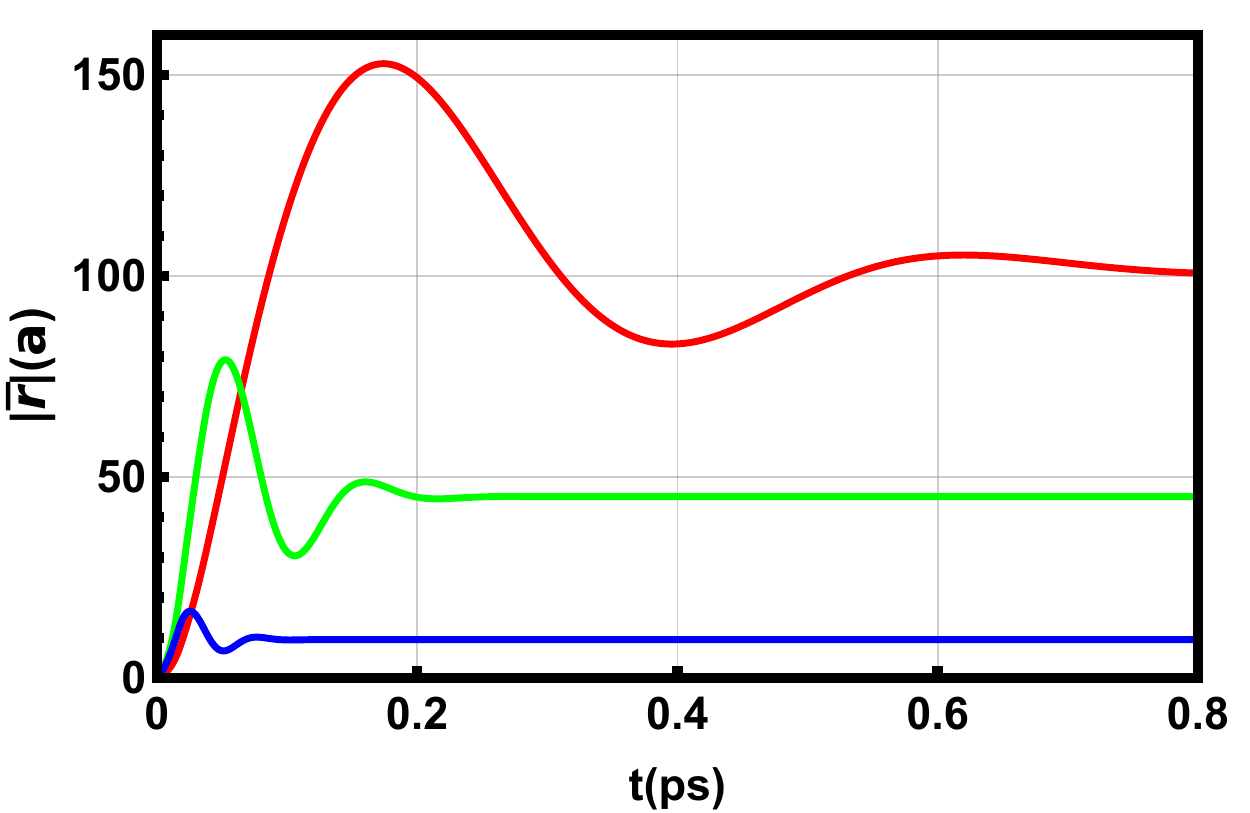}
        \label{FigrADPq20:SubFigC}
    }
    \caption{
        (color online) The average of position operator $|\bar r|$ versus the time near EDPs located at zero energy for $q_2=0$, $\mathbb{V}=7\pi$, $\theta_0=0$. \protect\subref{FigrADPq20:SubFigA}:  $\ell=1$, $\kappa_0=0.02~a^{-1}$ for different values of the  wave packet width $\sigma= 50~a^{-1}$ (red), $100~a^{-1}$ (green), $150~a^{-1}$ (blue), $200~a^{-1}$ (magenta). \protect\subref{FigrADPq20:SubFigB}: $\ell=1$, $\sigma= 150~a^{-1}$ for different values of $\kappa_0= 0.02~a^{-1}$ (red), $0.03~a^{-1}$ (green), $0.04~a^{-1}$ (blue), $0.05~a^{-1}$ (magenta). \protect\subref{FigrADPq20:SubFigC}:  $\sigma= 150~a^{-1}$, $\kappa_0=0.02~a^{-1}$ for different values of $\ell= \pm 1$ (red), $\pm 2$ (green), $\pm 3$ (blue).
    }
    \label{FigrADPq20}
\end{figure}

In addition of ODP, we have EDPs located at (${k_D}_x=0$, ${k_D}_y=\pm\sqrt{\mathbb{V}^2-(2\ell\pi)^2}/d$, $\varepsilon_D=0$) with $\ell\neq 0$ and $\mathbb{V}\geq 2\ell\pi$. To study ZB oscillations near these EDPs, we choose $\mathbb{V}=7\pi$, in order to have three EDPs at $\varepsilon=0$.
Figure \ref{FigrADPq20} %\subref{FigrADPq20:SubFigA} 
shows the ZB oscillations of the average of position operator $|\bar r|$ as function of the time 
%for some Gaussian wave packet 
near the first EDP at zero energy for $q_2=0$, $\mathbb{V}=7\pi$, $\theta_0=0$ and different values of the parameters $(\sigma, \kappa_0, l)$. 
In Figure \ref{FigrADPq20}\subref{FigrADPq20:SubFigA} with 
$\kappa_0=0.02~a^{-1}$ and $\ell=1$,
it is found that for small $\sigma$ there are no oscillations. Increasing $\sigma$, the amplitude of the ZB oscillations become larger compared to that of ODP and the attenuation becomes slow. In Figure \ref{FigrADPq20}\subref{FigrADPq20:SubFigB} with $\sigma= 150~a^{-1}$
 and $\ell=1$, we consider different values of the initial momentum $\kappa_0$ %with $\ell=1$, $q_2=0$, $\sigma=150~a$, and $\theta_0=0$. It is 
and show  that by decreasing  $\kappa_0$ the amplitudes of the ZB oscillations decrease, the frequency increases and the attenuation remains constant.
In Figure \ref{FigrADPq20}\subref{FigrADPq20:SubFigC} with $\sigma= 150~a^{-1}$ and $\kappa_0=0.02~a^{-1}$, for  different values of $\ell$ 
%with $q_2=0$, $\sigma=150~a$, and $\theta_0=0$. Note that 
we observe that by increasing $\ell$ the amplitudes, period and the attenuation decrease, 
%it is understandable, 
which is quiet normal because we have the relation 
\begin{equation}
    \dfrac{\omega}{\omega_0}=\dfrac{2\ell\pi}{\mathbb{V}}.
\end{equation}

%======================================================================

\begin{figure}[ht]\centering
    \subfloat[$\ell=1$]{
        \includegraphics[width=5.58cm]{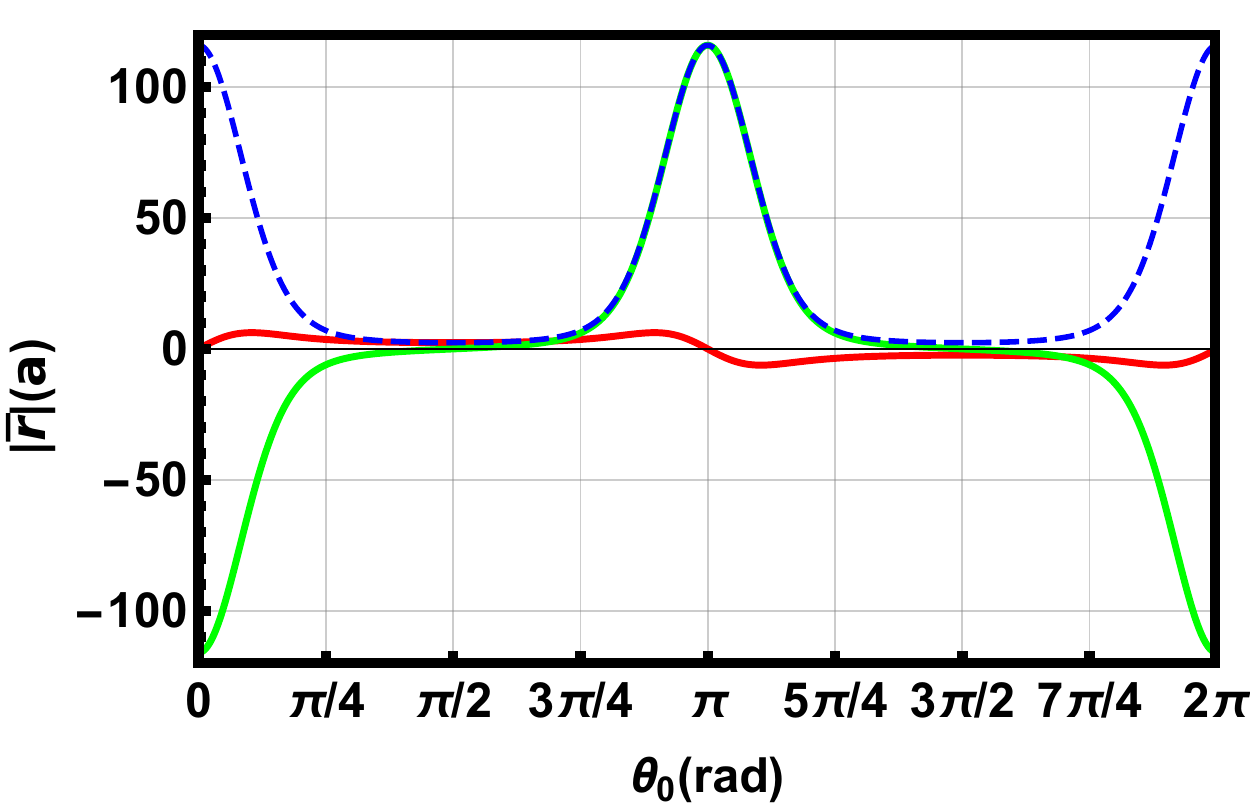}
        \label{FigrADPtheta:SubFigA}
    }
    \subfloat[$\ell=2$]{
        \includegraphics[width=5.58cm]{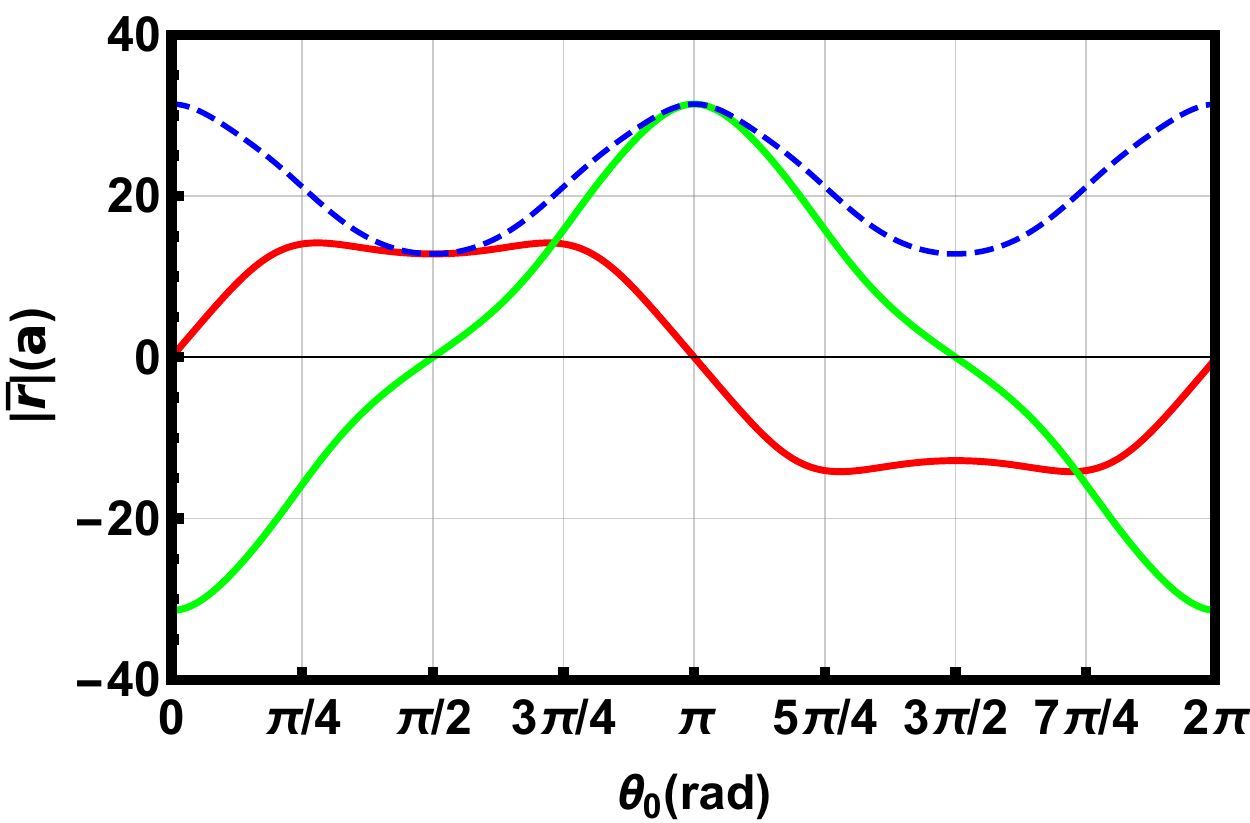}
        \label{FigrADPtheta:SubFigB}
    }
    \subfloat[$\ell=3$]{
        \includegraphics[width=5.58cm]{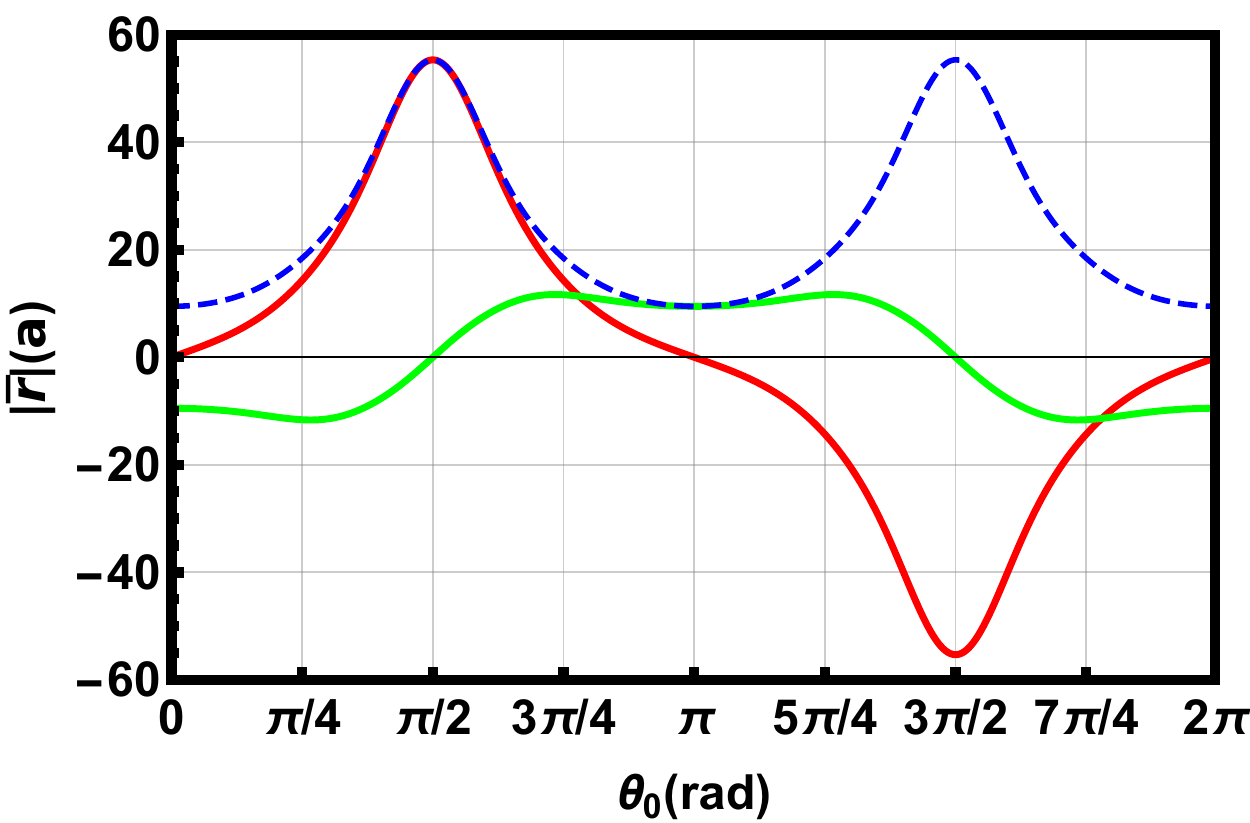}
        \label{FigrADPtheta:SubFigC}
    }
    \caption{
        (color online) The averages of position operators 
$\bar x_{11}$ (red), $\bar y_{11}$ (green) and $|\bar r|$ (blue) versus the angle $\theta_0$ near EDPs located at zero energy $\varepsilon_D=0$ for $t=0.1~\pico\second$, $q_2=0$, $\mathbb{V}=7\pi$, $\sigma= 150~a^{-1}$, $\kappa_0=0.02~a^{-1}$
        and three values of the quantum number $\ell=1,2,3$. 
        %$x$-direction, $y$-direction and radial direction ZB oscillation corresponding to red, green, and blue curves, respectively.
    }
    \label{FigrADPtheta}
\end{figure}

Figure $\ref{FigrADPtheta}$
presents  the averages of position operators 
$\bar x_{11}$ (red), $\bar y_{11}$ (green) and $|\bar r|$ (blue)  as  function of the angle $\theta_0$ for time $t= 0.1~$ps. It is clearly seen that for $\ell=1$, it shows that the $y$-direction ZB oscillation is predominate compared to the $x$-direction one. The amplitude of the 
$y$-direction ZB oscillation is very large in the vicinity of $\theta_0=0~[\pi]$. Now  for $\ell=2$, the amplitude of $x$-direction ZB oscillation increases but that of the $y$-direction decreases. Finally for $\ell=3$, the amplitude of the $x$-direction ZB oscillation is still increasing to reach large value in the vicinity of $\theta_0=\pi/2~[\pi]$.
In all cases, we observe that the behavior of $|\bar r|$ shows different periodicity with respect to $\theta_0=\pi/2$.

%=================================================================
\begin{figure}[!hbt]\centering
    \subfloat[$\mathbb{V}=3\pi$]{\centering
    \includegraphics[width=8.5cm]{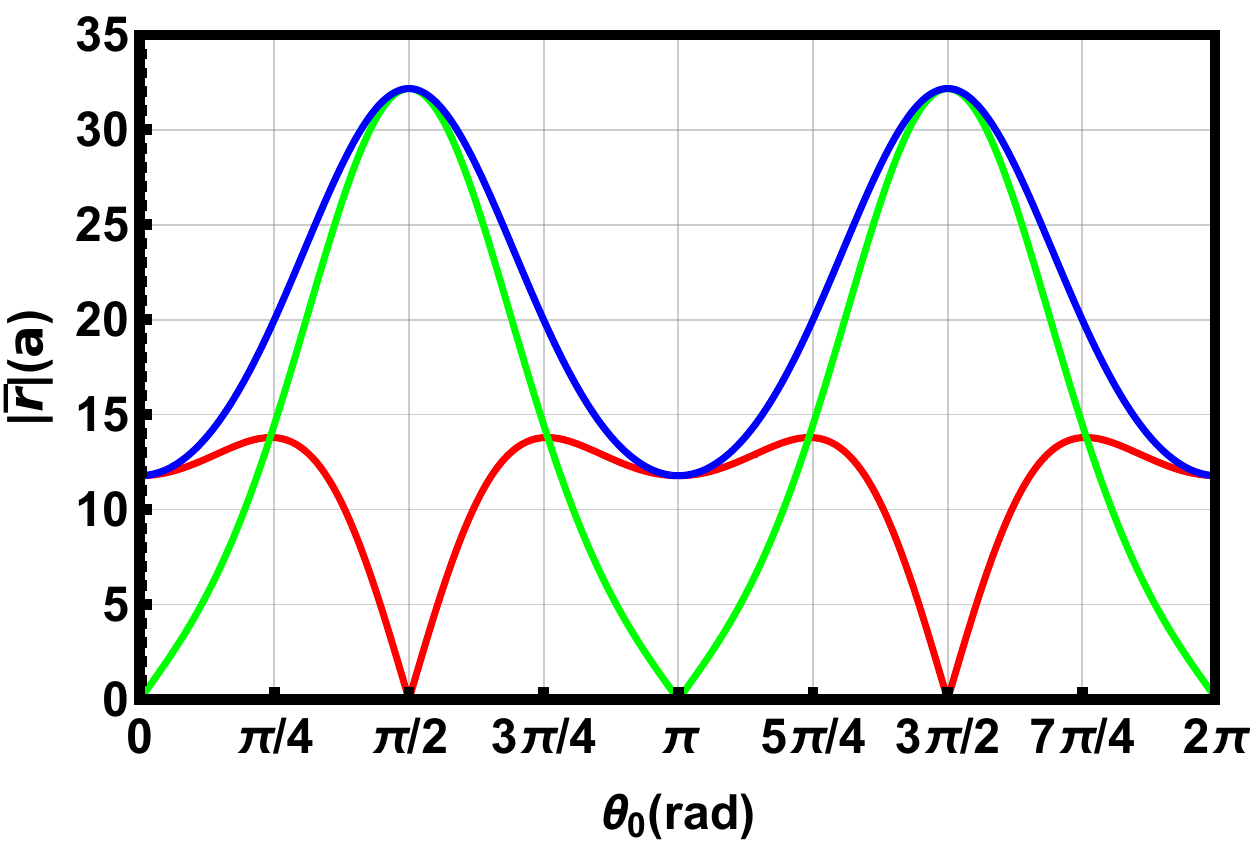}
    \label{rVDPmEventm2Theta}
    }
    \subfloat[$\mathbb{V}=2\pi$]{\centering
    \includegraphics[width=8.5cm]{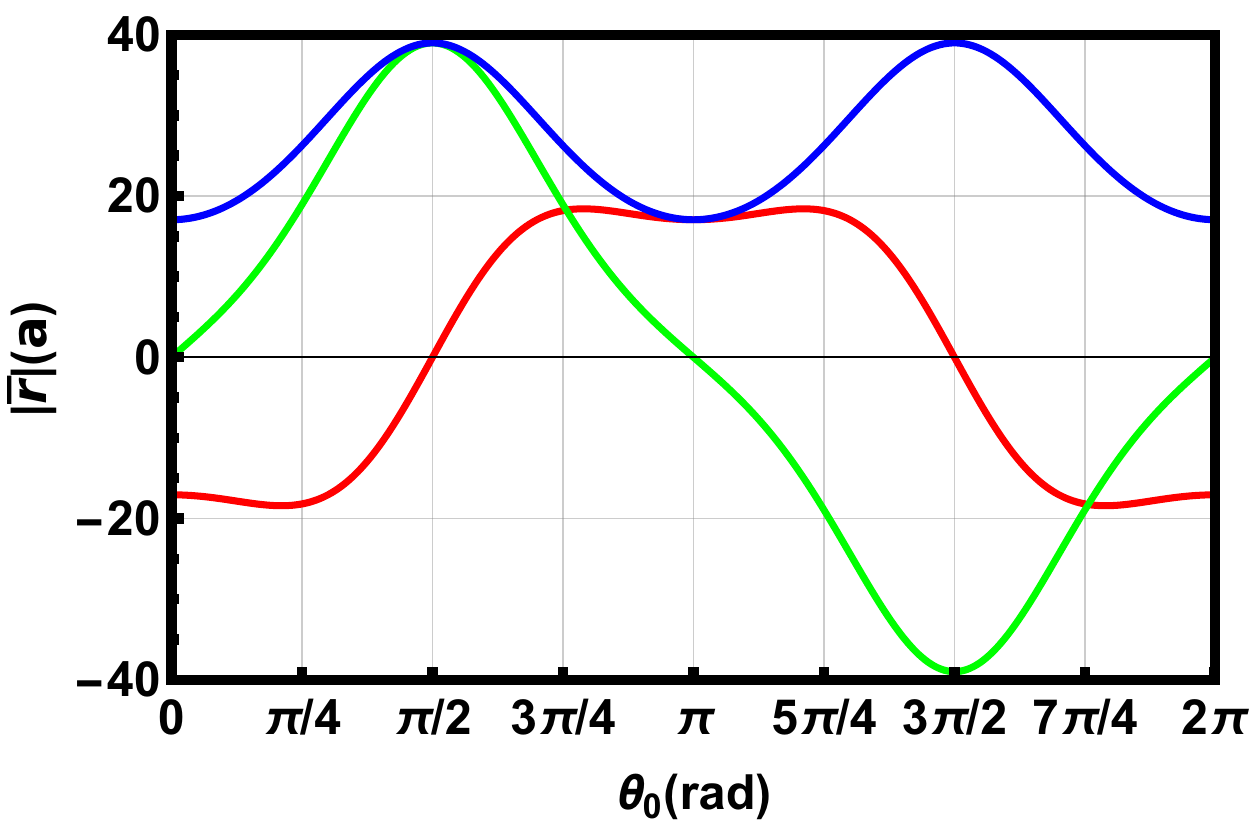}
    \label{rVDPmOddtm1Theta}
    }
    \caption{
    (color online) 
       The averages of position operators 
$\bar x_{11}$ (red), $\bar y_{11}$ (green) and $|\bar r|$ (blue) versus the angle $\theta_0$ for $t=14~\femto\second$, $q_2=1/3$, $\kappa_0=0.03~a^{-1}$. \protect\subref{rVDPmEventm2Theta}: near EDPs located at finite energy $\varepsilon_D=2\pi$ with ${k_D}_x={k_D}_y=0$. \protect\subref{rVDPmOddtm1Theta}: near EDPs located at finite energy $\varepsilon_D=\pi$ with ${k_D}_x=\pm\pi/d$, ${k_D}_y=0$. 
%The $x$ (red), $y$ (green) and radial (blue) directions of ZB oscillations.
    }
    \label{fig6}
\end{figure}

 In Figure \ref{fig6}, we present        the averages of position operators 
$\bar x_{11}$ (red), $\bar y_{11}$ (green) and  $|\bar r|$ (blue)
versus the angle $\theta_0$
near EDPs located at finite energy $\varepsilon_D=2\pi$ ($\varepsilon_D=\pi$)  for $t=14~\femto\second$, $\sigma=$, $q=1/3$, $\kappa_0=0.03~a^{-1}$.
%
%We study ZB oscillations near EDPs located at finite energy $\varepsilon_D=\pi m$ with ${k_D}_x=0$, ${k_D}_y=0$, $m\neq 0$. Figure \ref{fig6} present ZB oscillation versus angle $\theta_0$ near EDPs located at finite energy $\varepsilon_D=2\pi$ ($\varepsilon_D=\pi$)  for $t=14~\femto\second$. 
In the present  situation, we consider two interesting cases, for $\theta_0\neq\pi/2~[\pi]$
%it is interesting to stress that there are two  
we observe that   both $x$- and $y$-direction ZB oscillations exist but show different behaviors. For $\theta_0=\pi/2~[\pi]$, it appears that the amplitude of the $x$-direction ZB oscillations is null while that of $y$-direction becomes maximal.

\begin{figure}[!hbt]\centering
    \subfloat[$m=1, 5, 7, 11$]{
    \includegraphics[width=8.5cm]{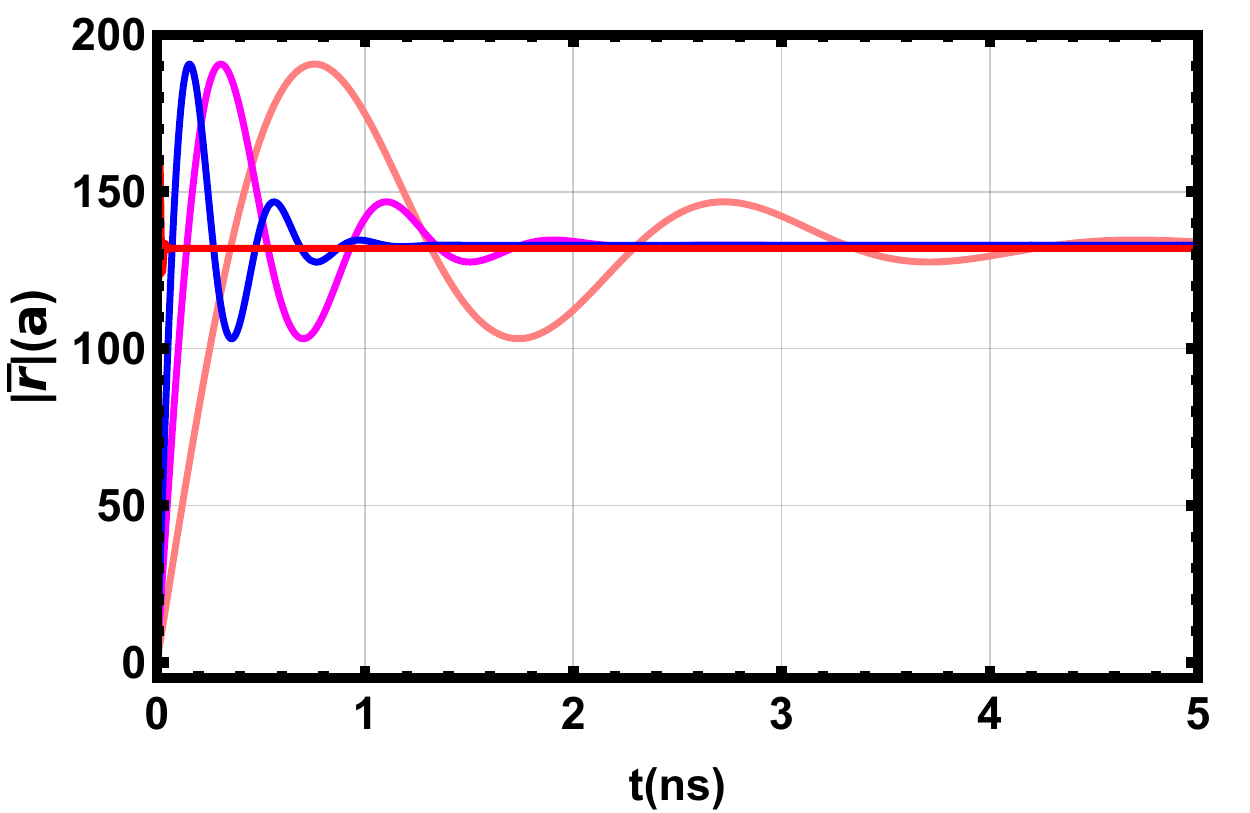}
    \label{FigrEDPmOddq13:SubFigA}
    }
    \subfloat[$m=3, 9$]{
        \includegraphics[width=8.5cm]{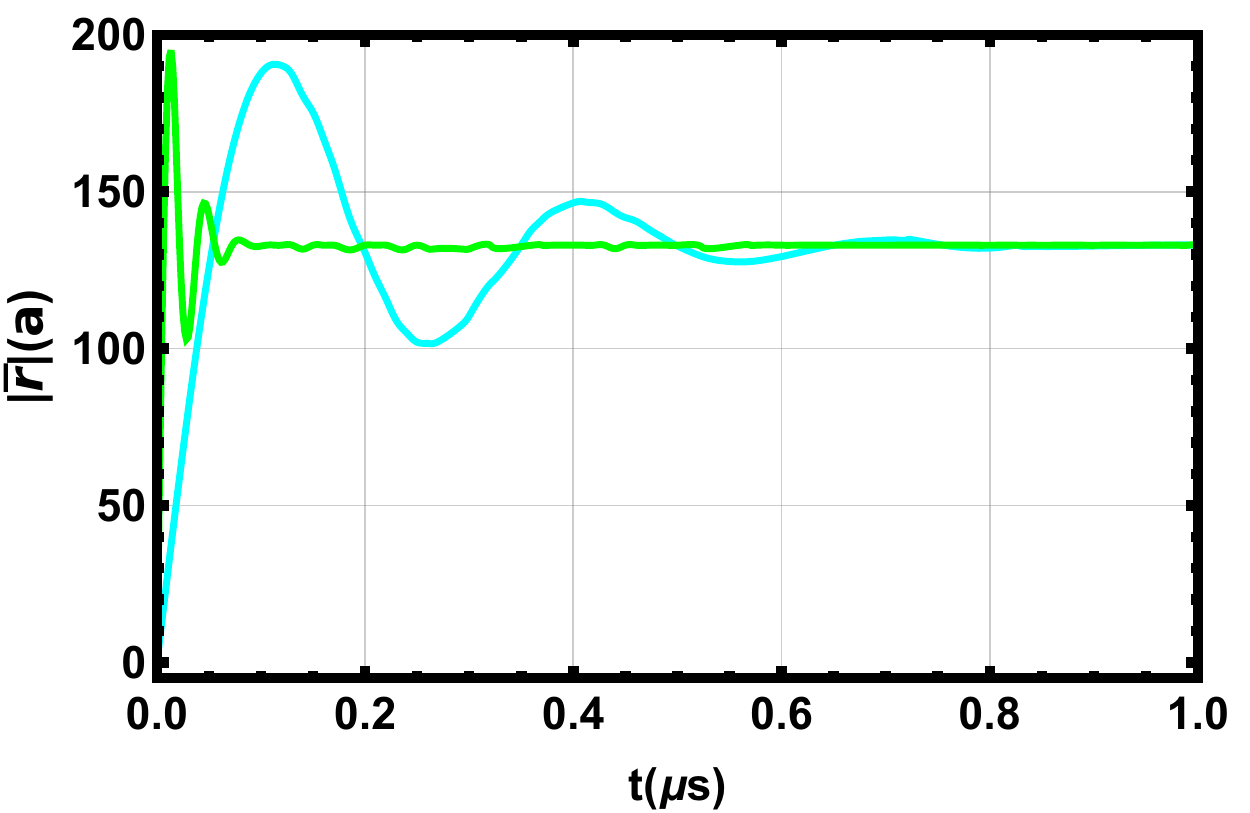}
        \label{FigrEDPmOddq13:SubFigB}
    }
    \caption{(color online) The average of position operator 
 $|\bar r|$ versus the time near EDPs located at finite energy energy $\varepsilon_D=\pi m$ with $m=2n\pm 1$  for $q_2=1/3$, $\mathbb{V}=0.01\pi$, $\theta_0=\pi/2$, $\kappa_0=0.03~a^{-1}$. \protect\subref{FigrEDPmOddq13:SubFigA}: $m=1$ (red), $5$ (blue), $7$ (magenta), $11$ (pink). \protect\subref{FigrEDPmOddq13:SubFigB}: $m=3$ (green), $9$ (cyan).
    }
    \label{FigrEDPmOddq13}
\end{figure}

\begin{figure}[!hbt]\centering
    \subfloat[$m=2, 4, 8, 10$]{
        \includegraphics[width=8.5cm]{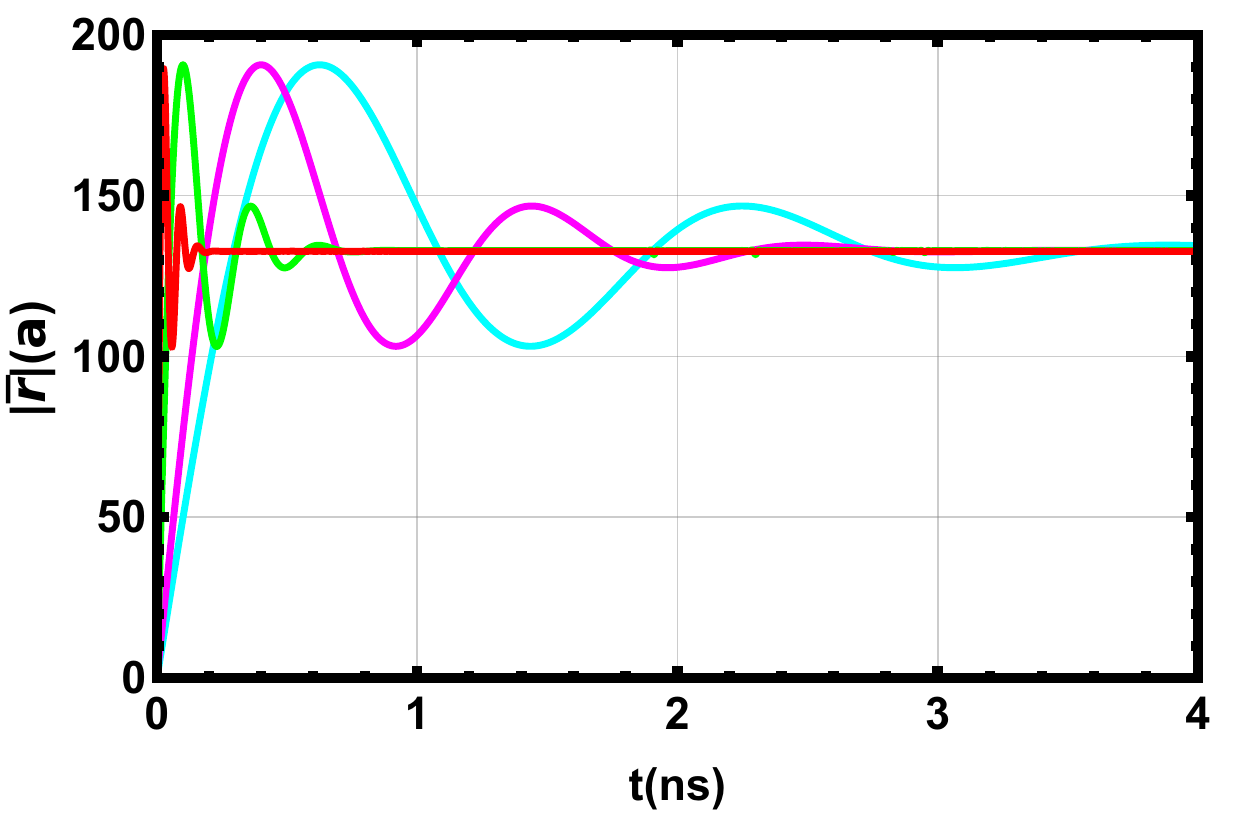}
        \label{FigrEDPmEvenq13:SubFigA}
    }
    \subfloat[$m=6, 12$]{
        \includegraphics[width=8.5cm]{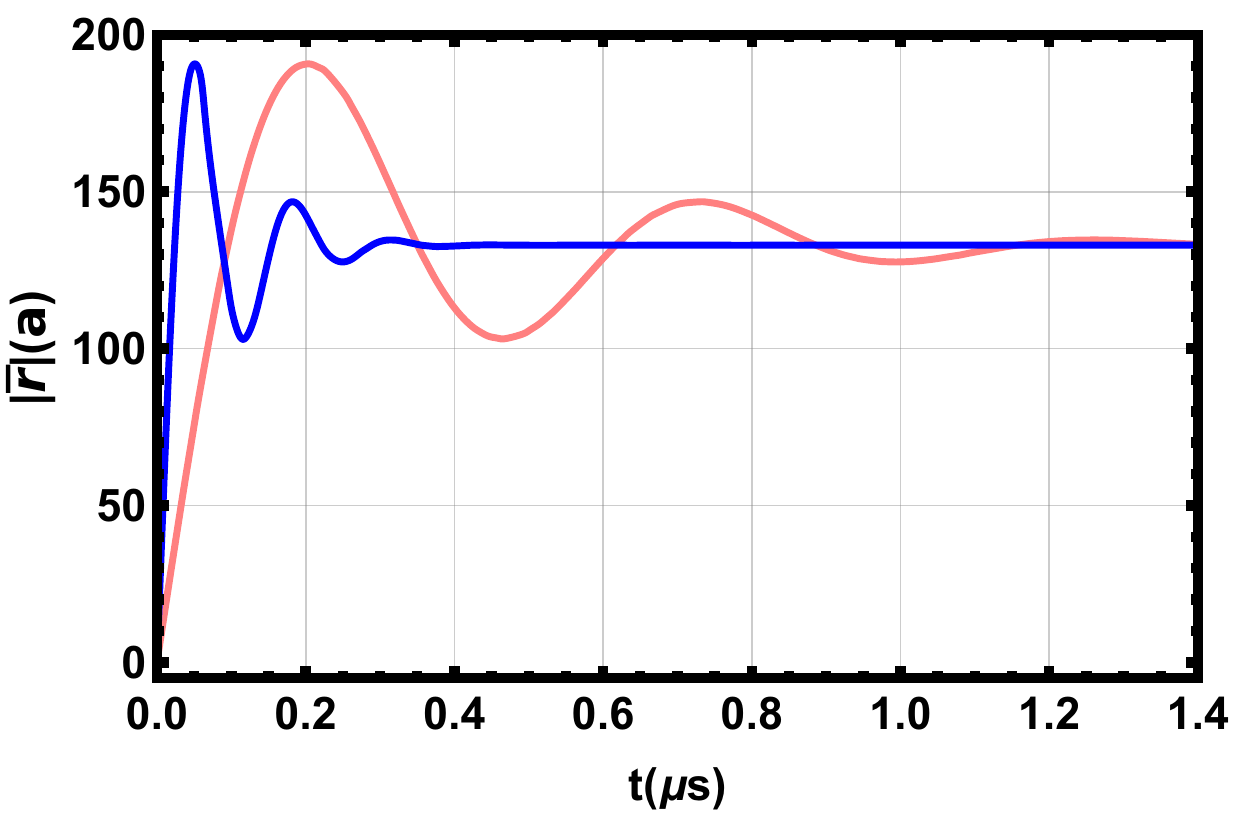}
        \label{FigrEDPmEvenq13:SubFigB}
    }
 \caption{
    (color online) The ZB oscillation versus time near EDPs located at finite energy energy $\varepsilon_D=\pi m$ with $m=2n\neq 0$  for $q_2=1/3$, $\mathbb{V}=0.01\pi$, $\theta_0=\pi/2$, $\kappa_0=0.03~a^{-1}$. \protect\subref{FigrEDPmEvenq13:SubFigA}: $m=2$ (red), $4$ (green), $8$ (magenta), $10$ (cyan). \protect\subref{FigrEDPmEvenq13:SubFigB}: $m=6$ (blue), $12$ (pink).
 }
 \label{FigrEDPmEvenq13}
\end{figure}

%Next we choose to study ZB oscillations for $\theta_0=\pi/2$ to be sure  
To get  maximal ZB oscillations in our system, we analyze the situations  where the angle takes the value $\theta_0=\pi/2$. Indeed, Figure \ref{FigrEDPmOddq13} (\ref{FigrEDPmEvenq13}) illustrates the ZB oscillations of the average of position operator $|\bar r|$ versus the time near EDPs located at finite  energy $\varepsilon_D=\pi m$ with $m=2n\pm 1$ ($m=2n\neq 0$) for some values of $m$. As long as $m$ increase, the frequency of ZB oscillations decreases except for $m=3$, $9$ ($m=6$, $12$). The period of ZB oscillations can reach a few nanoseconds, while the amplitude of oscillations remains almost constant and the frequency is much lower than that found in \cite{Wang2014PRA,Jianli2018JPCM}. However, the frequency decreases remarkably in the case of $m=3$, $9$ ($m=6$, $12$). To understand why there is such decrease in the frequency, it suffices to write the initial normalized frequency, using Table \ref{Tab:fxfy}, in terms of the normalized velocities
%under our conditions
\begin{equation}
\dfrac{\omega}{\omega_0}=
    \left\{
        \begin{array}{ccc}
            u_y^+ & \text{if}   & m=2n\pm 1\\
            u_y^- & \text{if}   & m=2n\neq 0.
        \end{array}
    \right.
\end{equation}

\begin{figure}[!hbt]\centering
    \subfloat[$m=2n\neq 0$]{
        \includegraphics[height=5cm]{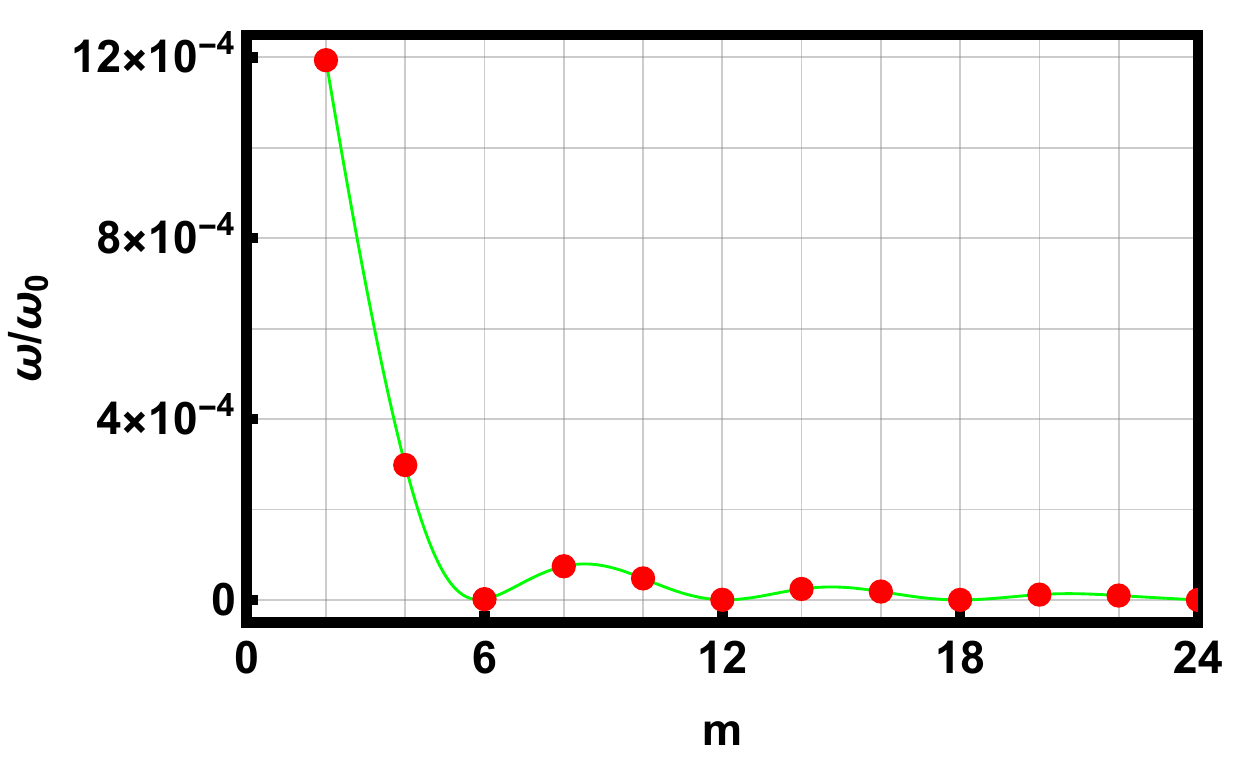}
        \label{VDPmEven_Omega}
    }
    \subfloat[$m=2n+1$]{
        \includegraphics[height=5cm]{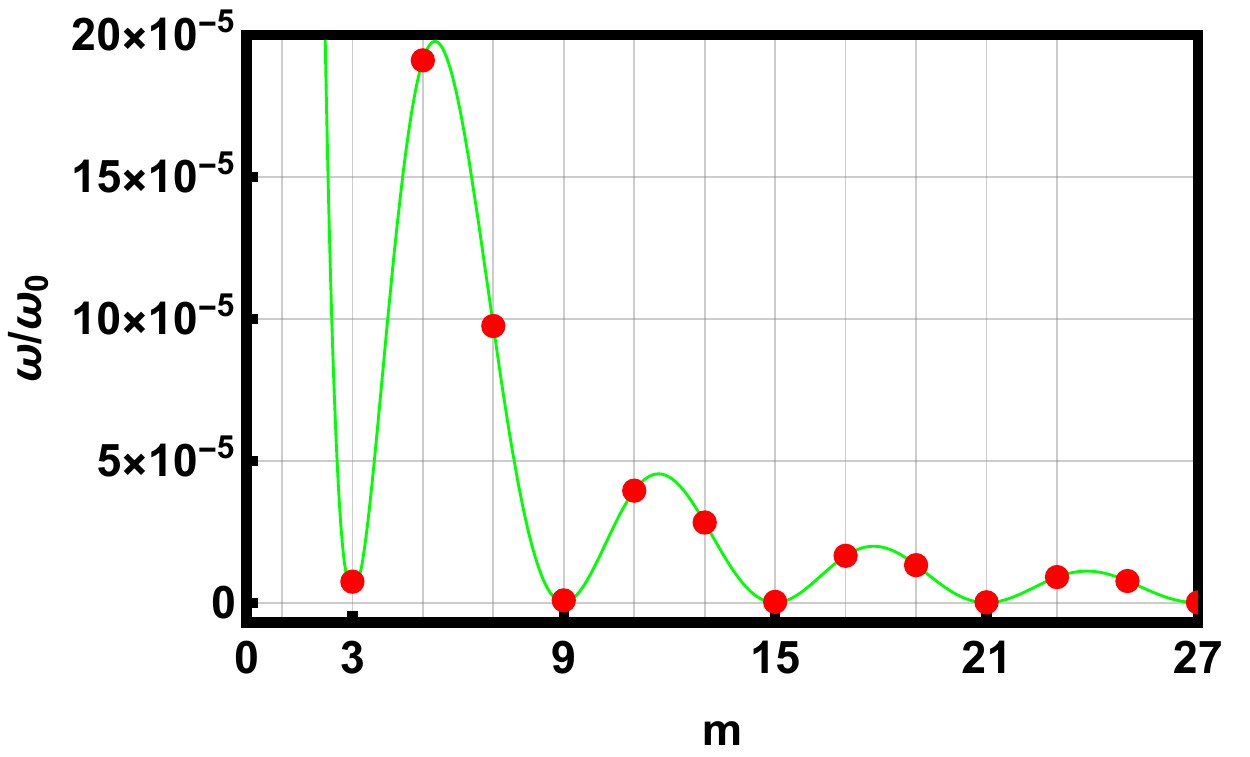}
        \label{VDPmOdd_Omega}
        }
    \caption{
        (color online) Initial normalized  frequency $\omega/\omega_0$ versus the quantum number $m$ for $\mathbb{V}=0.01\pi$ and $q_2=1/3$.
    }
    \label{figL555}
\end{figure}

Figure \ref{figL555} shows the initial normalized frequency $\omega/\omega_0$ versus the quantum number $m$ for $q_2=1/3$ and $\mathbb{V}=0.01\pi$. It is clearly seen that $\omega/\omega_0$ has the minima located at points $m=6k+1$ ($m=3(2k+1)$) with $k$ is integer.
For $m=6k$, the frequency reaches $10^8~\hertz$ and it is of order  $10^9~\hertz$, $10^8~\hertz$, $10^7~\hertz$  for $m=3,~9,~15$, respectively. While, the amplitude of the ZB oscillations  is of order $200~a$. This result suggest that  graphene with spacially modulated  may provide a good system to experimentally study the ZB effect near EDPs.

\begin{figure}[!hbt]\centering
    \subfloat[]{
        \includegraphics[height=5.23cm]{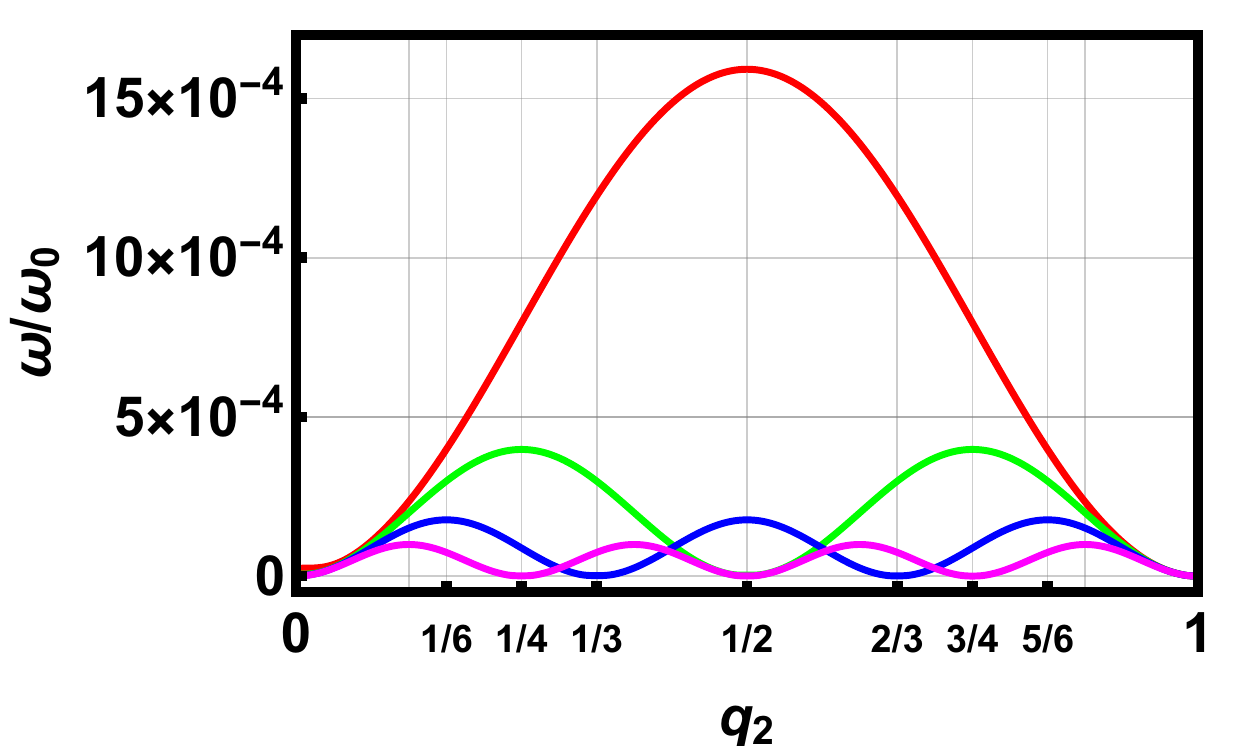}
        \label{FigOmegaVDPmEvenq2:SubFigA}
    }
    \subfloat[]{
        \includegraphics[height=5.23cm]{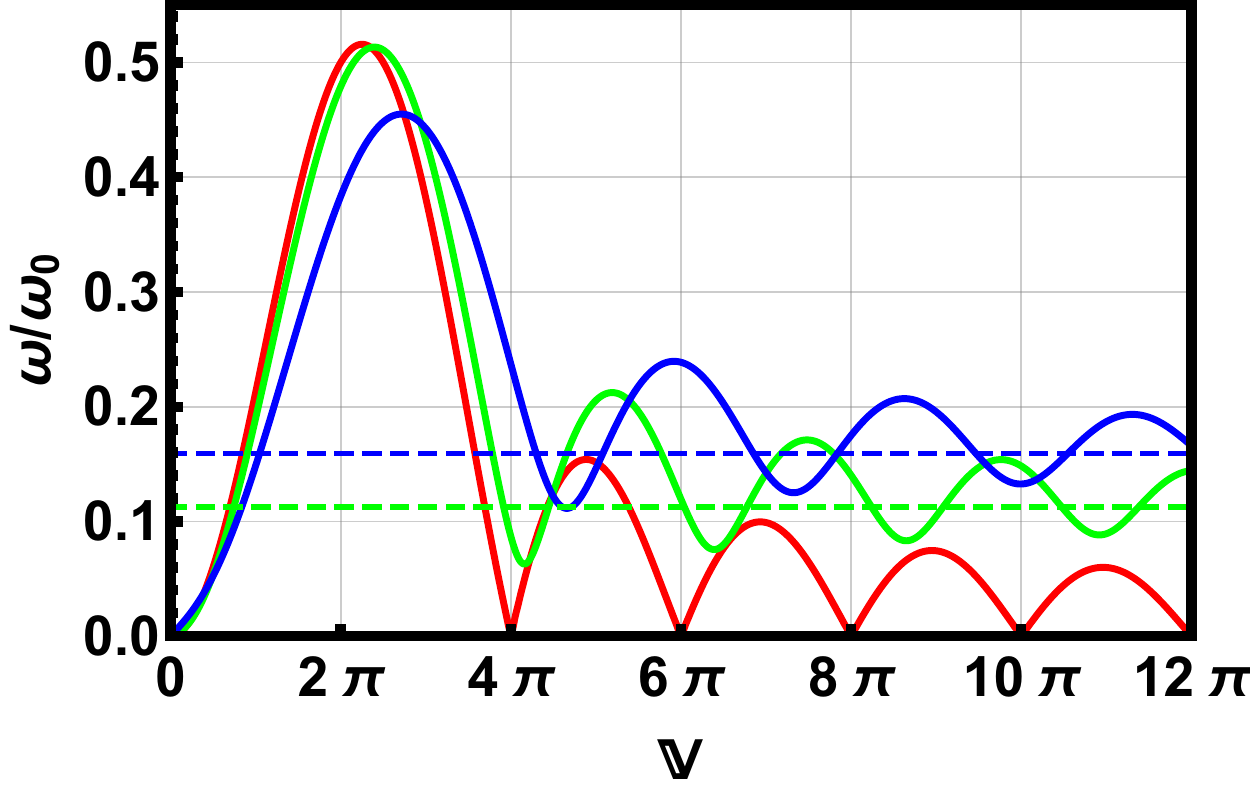}
        \label{FigOmegaVDPmEvenV:SubFigB}
    }
    \caption{
        (color online) The initial normalized  frequency $\omega/\omega_0$ such that \protect\subref{FigOmegaVDPmEvenq2:SubFigA}: versus the distance $q_2$ with $\mathbb{V}=0.01\pi$ for some values of $m=2$ (red), $4$ (green), $6$ (blue), $8$ (magenta), \protect\subref{FigOmegaVDPmEvenV:SubFigB}: versus the potential height $\mathbb{V}$ with $m=2$ for three values of the distance $q_2=0$ (red), $1/8$ (green), $1/4$ (blue).
    }
    \label{VDPmEvenq2}
 \end{figure}

\begin{figure}[!hbt]\centering
    \subfloat[]{
 		\includegraphics[height=5.23cm]{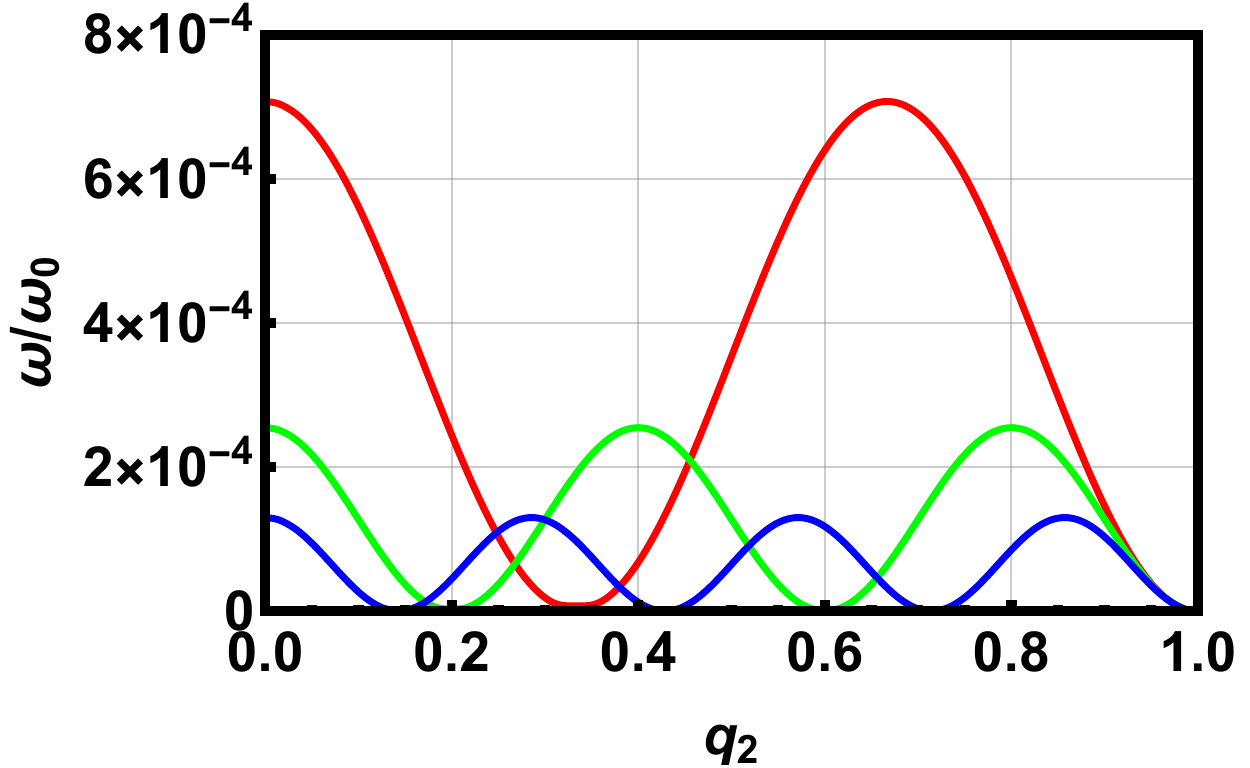}
 		\label{FigOmegaVDPmOddq2:SubFigA}
 	}
 	\subfloat[]{
 		\includegraphics[height=5.23cm]{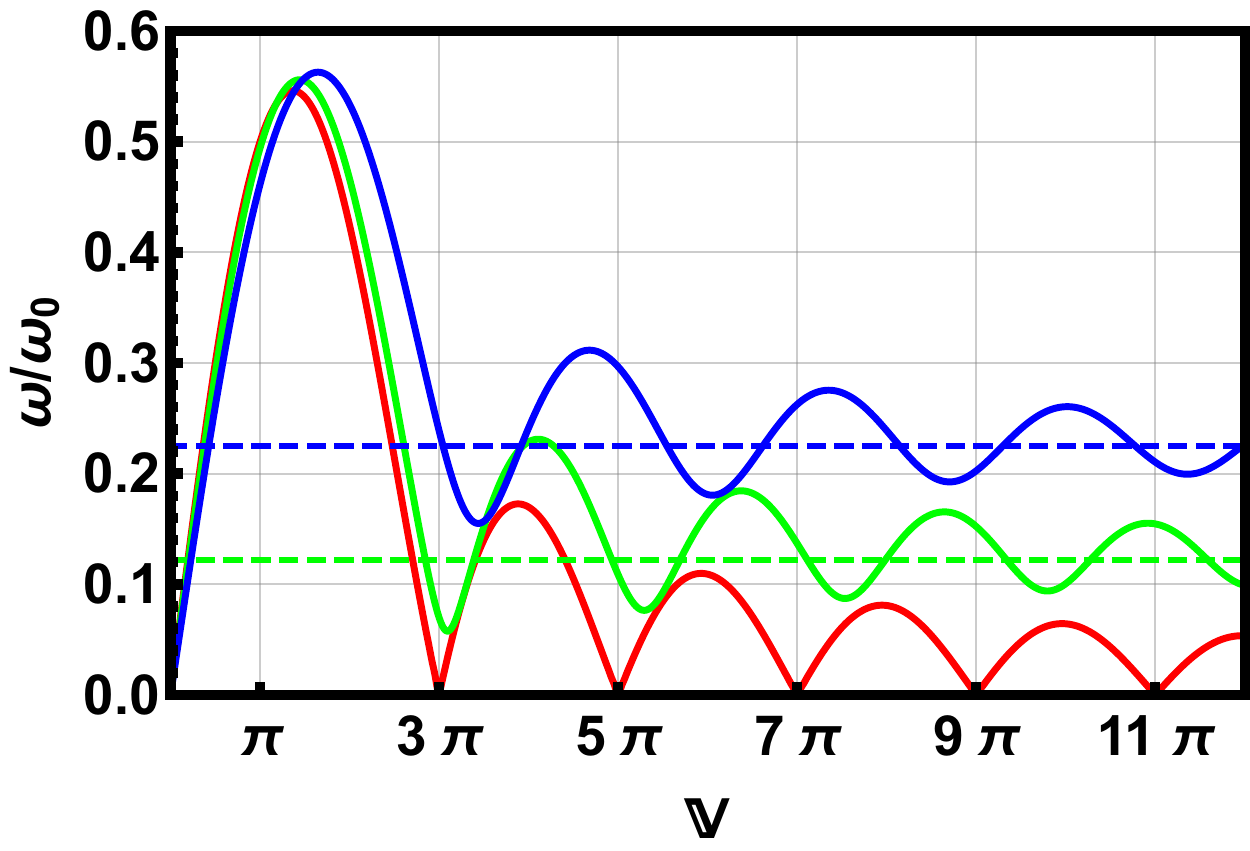}
 		\label{FigOmegaVDPmOddV:SubFigB}
 	}
  	\caption{
 		(color online) The initial normalized  frequency $\omega/\omega_0$ such that \protect\subref{FigOmegaVDPmOddq2:SubFigA}: versus the distance $q_2$ with $\mathbb{V}=0.01\pi$ for some values of $m=1$ (red), $3$ (green), $5$ (blue), $7$ (magenta).  \protect\subref{FigOmegaVDPmOddV:SubFigB}: versus the potential height $\mathbb{V}$ with $m=1$ for three values of the distance $q_2=0$ (red), $1/8$ (green), $1/4$ (blue).
   }
     \label{VDPmOddq2}
\end{figure}

Figures \ref{VDPmEvenq2}\subref{FigOmegaVDPmEvenq2:SubFigA} and \ref{VDPmOddq2}\subref{FigOmegaVDPmOddq2:SubFigA} elucidate the initial normalized frequency $\omega/\omega_0$ versus the distance $q_2$ with $\mathbb{V}=0.01$ for some values of $m$. As long as $m$ increases, $\omega/\omega_0$ decreases and the corresponding number of peaks increases. We notice that, for $m=2n\neq 0$,  $\omega/\omega_0$ has $n$ maxima  and $n-1$ minima. 
%The results of Figures \ref{VDPmEvenq2}\subref{FigOmegaVDPmEvenq2:SubFigA} and \ref{VDPmOddq2}\subref{FigOmegaVDPmOddq2:SubFigA} show  that the ZB oscillations of multi-unit graphene superlattice can be tunable by the distance $q_2$. 
In Figure \ref{VDPmEvenq2}\subref{FigOmegaVDPmEvenV:SubFigB} and \ref{VDPmOddq2}\subref{FigOmegaVDPmOddV:SubFigB}, we present the initial normalized frequency $\omega/\omega_0$ versus the potential height $\mathbb{V}$ with some values of $q_2$ for $m=2$. According to Figure  \ref{VDPmEvenq2}\subref{FigOmegaVDPmEvenV:SubFigB} (\ref{VDPmOddq2}\subref{FigOmegaVDPmOddV:SubFigB}), we summarize the following
interesting results. Indeed,
%\begin{itemize}
    %\item 
    when $q_2$ increases, the amplitude of oscillations of $\omega/\omega_0$ decreases and its periodicity increases as long as $\mathbb{V}$ increases. Up to a large value of $\mathbb{V}$, we obtain %end up with
    \begin{equation}
       \lim\limits_{\mathbb{V} \rightarrow +\infty} \dfrac{\omega}{\omega_0}= \dfrac{|\sin(q_2 m \pi )|}{m \pi}.
       \label{eq412}
    \end{equation}
    %\item 
    For $q_2=0$, we observe there are different oscillations
    %there is a succession of crests, which 
    appearing between each $2\pi$ and their amplitudes decrease when $\mathbb{V}$ increases except for the interval $[2\pi, 4\pi]$ $\left([\pi, 3\pi]\right)$. In addition, the values for $\omega/\omega_0=0$ %is null 
    coincide with those where EDPs appear, namely $\mathbb{V}=(k+1)\pi$ ($\mathbb{V}=(2k+1)\pi$) except for $\mathbb{V}=2\pi$ ($\mathbb{V}=\pi$) because for a given $m$ the potential height should be $\mathbb{V}\neq m\pi$.
    For a given $m$, the peak of $\omega/\omega_0$ is located at a potential height
    %value just 
    greater than $m\pi$.
%\end{itemize}

%=========================
\section*{Conclusion}\label{Conclusion}
%=========================
We have studied the Zitterbewegung (ZB) effect of massless Dirac fermions in  graphene
with spacially modulated potential near original Dirac point (ODP) extra Dirac points (EDPs). In the first Brillouin zone, we have seen that our system could influence the ZB effect by changing the group velocity of fermions. %It was found that the massless Dirac fermions %of the multi-unit graphene superlattice 
Such velocity is maximal along the superlattice direction and minimal along the perpendicular one. Subsequently,
using a Gaussian wave packet with finite momentum $\bm{\kappa_0}$, 
%which its direction described by the angle $\theta_0$, 
we have shown that the frequency of the ZB oscillations can be influenced by the applied potential  parameters such as the distance $q_2$, potential height $\mathbb{V}$, momentum $\kappa_0$, angle $\theta_0$ together with EDPs. 
%extra Dirac points type. 

The numerical calculations showed that to get  the large ZB oscillations, %we must put 
the wave packet center should be  at $\theta_0=0$ for EDPs located at zero-energy or at $\theta_0=\pi/2$ for ODP and EDPs located at finite energy $\varepsilon=m\pi$, with $m$ is integer. We have seen that the amplitude of the ZB oscillations can reach hundreds of angstroms
%We have found that, 
and their
frequency  can be in the interval $[10^{7}~\hertz$, $10^{9}~\hertz$]. We have shown that its attenuation can be slowly transient and can reach a few nanoseconds until microseconds, which can be clearly detected. Our results suggest that the  present system may provide an appropriate candidate  to experimentally realize the ZB effect near EDPs.

%=========================
\section*{Acknowledgment}
%=========================
The generous support provided by the Saudi Center for Theoretical Physics (SCTP) is highly appreciated by all authors.
%=================

%=================
\end{document}